\documentclass[11pt]{article}
\pdfoutput=1

\usepackage[utf8]{inputenc}
\usepackage[normalem]{ulem}
\usepackage{verbatim}
\usepackage{amsmath,amssymb,amsthm,amsfonts,mathrsfs,dsfont,braket,array,float}
\usepackage{color,graphicx,cite}
\usepackage{slashed}
\usepackage{enumitem}

\usepackage{epsfig}
\usepackage{mathabx}
\usepackage[mathscr]{eucal}
\usepackage{stmaryrd}
\usepackage{esint}
\usepackage{physics}
\usepackage{setspace}
\usepackage{braket}
\usepackage{float}
\usepackage{url}
\usepackage{mathtools}
\usepackage{bbold}
\usepackage{slashed}
\usepackage{tikz}
\usepackage{graphicx}
\usepackage{epstopdf}
\usepackage{subfigure}
\usepackage[labelfont=bf]{caption}
\usepackage{pgfplots}
\usepackage{fancybox}
\usepackage{eurosym}
\usepackage{tcolorbox}
\usepackage{tensor}
\allowdisplaybreaks

\usepackage[margin = 2.2cm]{geometry}
\usepackage[ragged]{footmisc}
\usepackage[bookmarks=true,bookmarksnumbered=false,
hyperindex=true,bookmarksopen=true,hyperfigures=true,
colorlinks=true,linkcolor=darkgray,citecolor=custom,urlcolor=custom,breaklinks]{hyperref}
\definecolor{webgreen}{rgb}{0, 0.5, 0}
\definecolor{webblue}{rgb}{0.06, 0.2, .65}
\definecolor{webblue2}{rgb}{0, 0.33, .71}
\definecolor{webred}{rgb}{0.9, 0.1, 0}
\definecolor{darkgreen}{rgb}{0,0.4,0.2}
\definecolor{custom}{rgb}{0.05,0.31,0.55}
\definecolor{darkgray}{rgb}{0.4,0.4,0.4}

\def\be{\begin{equation}}
\def\ee{\end{equation}}

\numberwithin{equation}{section}

\def\la{\left\langle}
\def\ra{\right\rangle}

\begin{document}
	\setcounter{secnumdepth}{2}
	
	\begin{titlepage}
		\leftline{}%{\timestamp}
		\vskip .25cm
		\hspace{13.5cm}	BRX-TH-6711
		\vskip 2cm
		%\centerline{\LARGE \bf Aspects of wormholes and heavy operator statistics}
		\centerline{\LARGE \bf Wormholes from heavy operator statistics}
		\vskip .4cm
		\centerline{\LARGE \bf in AdS/CFT}
		%\centerline{\LARGE \bf Deriving a wormhole in AdS/CFT}
		
		\bigskip
		
		\vskip 1.cm
		
		\centerline{\large Martin Sasieta}

		\vskip 1.25cm
		%\centerline{\bf}

		\centerline{\sl  Martin Fisher School of Physics, Brandeis University,}
		\centerline{\sl  Waltham, Massachusetts 02453, USA}
		
		\vskip 0.4cm
		
		\centerline{\tt martinsasieta@brandeis.edu}

		\vskip 2.5cm
		\centerline{\bf Abstract} \vskip 0.2cm \noindent

			We construct higher dimensional Euclidean AdS wormhole solutions that reproduce the statistical description of the correlation functions of an ensemble of heavy CFT operators. We consider an operator which effectively backreacts on the geometry in the form of a thin shell of dust particles. Assuming dynamical chaos in the form of the ETH ansatz, we demonstrate that the semiclassical path integral provides an effective statistical description of the microscopic features of the thin shell operator in the CFT. The Euclidean wormhole solutions provide microcanonical saddlepoint contributions to the cumulants of the correlation functions over the ensemble of operators. We finally elaborate on the role of these wormholes in the context of non-perturbative violations of bulk global symmetries in AdS/CFT.
		
		\noindent
		
	\end{titlepage}
	%\small
	
	{\hypersetup{linkcolor=black}
	\tableofcontents}
	
	\vspace{.5cm} \hrule

	\vspace{.5cm}
	
	\section{Introduction}
	
	%{\color{red} Wormholes in general, then ETH and wormholes, then main results of the paper, then organization of the paper}
	
	One of the outstanding aspects of gravity is the apparently simple characterization that some of its non-perturbative features admit in terms of spacetime geometry. But perhaps more remarkable is the fact that some of these features can be reproduced rather directly in the semiclassical quantization of gravity, once `spacetime instantons' are included into the oracle of the path integral \cite{GibbonsHawking,GibbonsHawking2,HP}. 
	
	Altogether, the semiclassical prescription consists in fixing boundary conditions in a region $M$, typically of weak gravity, and considering all possible topologically distinct solutions $X$ to the gravitational equations, respecting the boundary conditions $\partial X = M$, in a formal asymptotic expansion of the form
	\be\label{eq:semiclassicalpi}
	 Z_{\text{grav}}[M] \sim \sum_{X}\, e^{-I[X]}\, Z_X\,.
	 \ee
	 Each spacetime configuration $X$ is weighted by its classical gravitational action $I[X]$, and it includes the term $Z_X$ representing the path integral of the quantum fields, or ideally of the underlying string theory, expanded around the background manifold $X$.  
	 
	%The avatar of this phenomenon is the black hole entropy, given in terms of the area of the horizon in Planck units. 

	The predictions of the semiclassical rules of \eqref{eq:semiclassicalpi} can be put under scrutiny in gravitational systems for which the non-perturbative description is under control. Namely, these include the handful of models of string/M-theory in AdS space with known CFT duals. In these models, the holographic dictionary naturally identifies the semiclassical path integral of the bulk supergravity, $Z_{\text{grav}}[M]$, where $M$ is now the asymptotic boundary of AdS, with the path integral of the holographic system $Z_{\text{CFT}}[M]$ defined in the rigid conformal boundary $M$ of AdS \cite{Witten:1998qj}. The latter is defined in the large-$N$ asymptotic expansion, which corresponds to the semiclassical expansion in the bulk,
	\be\label{eq:semiclassicalpiholo}
	 Z_{\text{CFT}}[M] \sim Z_{\text{grav}}[M]\,.
	 \ee
	With this identification in mind, the semiclassical path integral provides a bulk mechanism that successfully accounts for many non-perturbative properties of the holographic system, such as: the large-$N$ confinement/deconfinement phase transition \cite{HP,Witten:1998qj, Witten:1998zw}, the quantum noise in the thermal correlation functions at late times \cite{Maldacena:2001kr, Barbon:2003aq,Barbon:2014rma} or the entanglement structure of holographic states, once the boundary replica trick is complemented with these rules \cite{Headrick:2010zt,LM,FLM,Dong:2016fnf}.

 	Despite its obvious success, the semiclassical prescription comes equipped with extra configurations, which seem to be \textit{a priori} admissible according to the low-energy effective rules of \eqref{eq:semiclassicalpi}, and which spoil the harmony with the holographic description. These are the so-called `Euclidean wormhole' solutions, connected geometries $X$ which possess multiple disconnected asymptotic boundaries $\partial X = M_1 \cup ... \cup M_k$. Euclidean wormholes can be constructed within low-energy effective actions arising from string theory in AdS space \cite{Maldacena:2004rf,Arkani-Hamed:2007cpn,Marolf:2021kjc}. 
 	
 	When Euclidean wormholes are included, the semiclassical path integral $Z_{\text{grav}}[M_1\cup M_2]$ defined in two disconnected asymptotic boundaries no longer factorizes
 	\be\label{eq:fact}
	 Z_{\text{grav}}[M_1\cup M_2] \neq Z_{\text{grav}}[M_1]Z_{\text{grav}}[M_2]\,.
	 \ee
 in clear contradiction with the the manifest factorization of the path integral of two independent holographic systems. 
 
 This raises the problem of identifying the sense in which \eqref{eq:semiclassicalpiholo} holds only as an approximate relation. Of course, the lack of a proper derivation of the semiclassical rules \eqref{eq:semiclassicalpi} from underlying principles, which may ultimately exist within Euclidean string theory, makes it still possible that these rules require of substantial modification, and this could particularly affect wormhole solutions. However, the possibility of excluding wormholes seems unnatural from the low-energy viewpoint, given the simplicity and the proven effectiveness of the only rule of summing over all possible solutions.

 	In recent years, the role of wormholes has been revisited, driven by the substantial progress in the two-dimensional model  of Jackiw–Teitelboim dilaton-gravity. The full-fledged gravitational path integral of JT gravity, defined as a sum over topologies and volume integral over moduli spaces of Riemann surfaces, happens to actually provide a statistical description over an ensemble of quantum mechanical Hamiltonians \cite{Saad:2019lba}. In the spirit of random matrix universality, the ensemble is expected to approximate some of the properties of the chaotic Hamiltonian of the higher-dimensional black hole described by this model. Thus, in this case, the off-shell analog of \eqref{eq:semiclassicalpi} in two dimensions implements an explicit disorder average over theories, which provides a statistical approximation to the microscopic description of the system, i.e., the putative left-hand side in a relation of the form \eqref{eq:semiclassicalpiholo}.
 	
 	Semiclassical wormholes also play a prominent role in JT gravity. An example is the microcanonical double-cone geometry \cite{Saad:2018bqo},  a wormhole which accounts for the characteristic late-time `ramp' in the spectral correlation functions of typical members of the ensemble of Hamiltonians, a regime which is dominated by the statistics of nearby eigenvalues \cite{Cotler:2016fpe}.  A somewhat simpler class of wormhole solution exists when the bulk theory is coupled to matter particles. These wormholes become stable as a consequence of the backreaction of the matter particles, and they contribute to the statistical description of the correlation functions of the dual operators \cite{Stanford}.

 	In any case, one can gather all this evidence in low-dimensional models in order to seek for an independent interpretation of the semiclassical path integral \eqref{eq:semiclassicalpi} in higher dimensional AdS/CFT. The natural guess is that the low-energy rules of \eqref{eq:semiclassicalpi} actually provide an effective statistical description of the chaotic properties of the holographic system, in particular of those properties which are associated to black hole microstates \cite{Saad:2019lba,Stanford}. The statistical description is able to faithfully compute coarse-grained (and yet non-perturbative) quantities of the gravitational system such as the number of states of the black hole, while, on the other hand, it fails to incorporate finer properties such as factorization. 
 	
 	The specific proposals so far identify the semiclassical prescription \eqref{eq:semiclassicalpi} with a statistical characterization of the CFT data in \eqref{eq:semiclassicalpiholo}, which includes the chaotic spectral properties of the CFT Hamiltonian \cite{Altland:2020ccq,Cotler:2021cqa} as well as the fine-grained structure of the OPE coefficients \cite{Belin:2020hea,Belin:2020jxr,Belin:2021ryy,Belin:2021ibv,Anous:2021caj,Chandra:2022bqq}. \footnote{Replica wormholes would also fit in this category, providing the approximate value for the entanglement spectrum of the radiation of an evaporating black hole \cite{Penington:2019kki,Liu, Sasieta:2021pzj}.} In a broader sense, semiclassical physics is expected to generate a statistical description of `simple' observables $\mathcal{O}$ in the chaotic high-energy spectrum of the CFT. The structure such a coarse-grained description is already implicit in the \textit{Eigenstate Thermalization Hypothesis} (ETH) \cite{Deutsch,Srednicki1994,Srednicki1999} for any chaotic system. Namely, the matrix elements of $\mathcal{O}$ in the energy basis will have the form
 \be\label{eq:ETH}
 \bra{E_n} \mathcal{O} \ket{E_m} \,=\,e^{-S(\bar{E})/2}\,g(\bar{E},\omega)^{1/2}\,R_{nm}\,,
 \ee
 for $\bar{E} = (E_n+E_m)/2$ and $\omega = E_n-E_m$. The matrix elements are exponentially suppressed in the microcanonical entropy $S(\bar{E})$, with a smooth $O(1)$ envelope function $g(\bar{E},\omega)$ which encodes the information about the microcanonical two-point function of the operator on each individual matrix element. In the form of \eqref{eq:ETH} we have neglected the smooth diagonal part present in the general ETH ansatz, by assuming that the simple operator in question has no $O(1)$ microcanonical trace. 
 
 The coefficients $R_{nm}$ in \eqref{eq:ETH} generally correspond to erratic $O(1)$ complex numbers that depend on the particular operator $\mathcal{O}$. The coarse-graining implicit in the ETH ansatz is to view these coefficients as a set of actual independent complex random variables of zero mean and unit variance, which parametrize an ensemble of operators. Typical members of this ensemble approximately share the microcanonical correlation functions with the original operator, and they only differ in the fine-grained structual phases between chaotic eigenstates. This is, of course, a mathematical trick that permits to make general statements. It is the reason why the ETH ansatz turns out to be so effective when proving local thermalization for generic quantum many-body systems, without the need to solve for the particular Hamiltonian of the system.
	
	In this paper, we examine the role of the statistical coarse-graining of simple operators in relation to the semiclassical description of holographic systems \cite{Saad:2019pqd,Pollack, Altland:2021rqn,Jafferis:2022uhu,Jafferis:2022wez}. For that purpose, we consider a heavy operator $\mathcal{O}$ which creates a spherical thin shell of dust particles and backreacts on the geometry. We assume that the shell operator $\mathcal{O}$ is simple enough in the internal space of the CFT to admit an ETH form. This is reasonable since $\mathcal{O}$ is constructed out of a product of approximate local single-trace operators at different points, which create a collection particles in the asymptotic region of AdS.
	
	We first extract the envelope function $g(\bar{E},\omega)$ appearing in the ansatz \eqref{eq:ETH} for the CFT operator $\mathcal{O}$, which controls the typical magnitude of the matrix elements of the operator over the microcanonical window, from the bulk thermal two-point correlation function of the thin shell, computed semiclassically. Since the operator is heavy, the envelope function $g(\bar{E},\omega)$ in this case scales in a particular way with the microcanonical entropy $S(\bar{E})$. This induces a conceptual modification with respect to the standard form of the ETH ansatz. Namely, the operator now is able to modify the dominant microcanonical contribution to thermal correlation functions in the thermodynamic limit. We show that this phenomenon is dual to the classical backreaction that the thin shell exerts on the spacetime geometry.
 
 We then proceed to define an actual ensemble of microscopic operators consistent with the effective thin shell description in the bulk, by fixing the value of the envelope function $g(\bar{E},\omega)$ and by promoting the coefficients $R_{nm}$ in the ansatz \eqref{eq:ETH} to actual random variables of zero mean and unit variance. We show that the cumulants of the thermal correlation functions over this ensemble of operators are captured by semiclassical wormhole contributions, associated to different microcanonical saddlepoint contributions of the CFT. Schematically, we show that the variance of the correlation functions is reproduced by the renormalized gravitational action $\Delta I[X_k]$ of an Euclidean wormhole solution $X_k$,
	\be\label{eq:introinterpwh}
\overline{ \langle \mathcal{O}_1... \,\mathcal{O}_k\rangle_\beta \;\langle \mathcal{O}_1... \,\mathcal{O}_k\rangle_\beta^*} \,\sim \,e^{-\Delta I[X_k]}\;.
	\ee
 We provide a generalization of this construction for multi-boundary Euclidean wormholes reproducing higher cumulants of the ensemble of microscopic operators.

	Based on previous discussions in the literature, we also analyze the role of these wormholes in the context of non-perturbative bulk global symmetry violating effects in higher dimensional AdS/CFT. We argue that the on-shell action of the wormhole \eqref{eq:introinterpwh} for $k=1$ provides the typical value of a non-perturbative violating thermal amplitude of any bulk global symmetry under which the thin shell is charged. We analyze the temperature-dependence of this amplitude from the classical wormhole action.

	The paper is organized as follows: In section \ref{sec:2} we propose an ETH ansatz for the matrix elements of the heavy shell operator in the energy basis. Then, we match the envelope function in this ansatz with the physics of the bulk thermal two-point function. In section \ref{sec:3} we use this ansatz to define an ensemble of heavy operators consistent with the semiclassical description of the thin shell. We then compute the connected contributions to products of thermal correlation functions over the ensemble, and match such contributions to the action of classical wormholes stabilized by thin shells. In section \ref{sec:4} we revisit previous arguments on typical global symmetry violating amplitudes induced by these wormholes. We end with some conclusions and some technical details in appendix \ref{appendix:A}.
	
	\section{Heavy shell operator}
	\label{sec:2}
	
	We consider a holographic CFT placed on a spatial $\mathbf{S}^{d-1}$ of radius $\ell$, and an operator $\mathcal{O}$ whose action is to inject a thin cloud of dust particles in the bulk, close to the asymptotic boundary of AdS. The operator is represented formally as
	\be\label{eq:operator}
	{\cal O}\; =\, \prod_{i=1}^n \,\phi(r_\infty, \theta_i) \; .
	\ee
	The effect of each insertion $\phi(r_\infty, \theta_i)$ is to create a massive dust particle in the bulk, centered at some fixed angular position $\theta_i \in \mathbf{S}^{d-1}$ within its Compton wavelength $\lambda \ll \ell$, at a bulk proper radius of order $r_\infty \sim \ell^2/\varepsilon$. In the CFT, the operator $\phi(r_\infty, \theta_i)$ is non-local over a domain $D^i_\varepsilon \subset \mathbf{S}^{d-1}$ of volume $\varepsilon^{d-1}$ associated with the support of its HKLL representation in terms of single-trace boundary operators \cite{Hamilton:2006az}. 
	
	We demand that the operator \eqref{eq:operator} is heavy, composed of a number of dust particles which scales parametrically with the central charge, $n\sim N^2 \sim G^{-1}$, where we adopt $\ell =1$ to set AdS units for the rest of the paper. Moreover, the particle insertions are approximately homogeneously distributed along the sphere. Thus, with this choice, the operator $\mathcal{O}$ creates an approximately spherical thin cloud of dust particles, which is heavy enough to classically backreact on the geometry, at leading order in the semiclassical $G\rightarrow 0$ expansion. In the bulk, the cloud can be effectively described as a presureless perfect fluid localized at the worldvolume $\mathscr{W}$ of a thin shell, with energy-momentum tensor
	\be
	T_{\mu\nu}\Big|_\mathscr{W} = \sigma u_\mu u_\nu \;,
	\ee
	where $\sigma$ is the surface density, and $u^\mu$ is the proper velocity of the fluid, tangent to $\mathscr{W}$. The spherical shell has a total rest mass
	\be
	m =  \sigma \,V_\Omega\, r_\infty^{d-1}\,,
	\ee
	where $V_\Omega = \text{Vol}(\mathbf{S}^{d-1})$. The effecrtive hydrodynamic description of the operator will be good enough as long as the fluid density $\sigma$ remains large compared to one dust particle per unit volume, and yet small enough in Planck units so that its classical description can be trusted.

\subsection{ETH ansatz}

In order to provide a full microscopic characterization of the operator $\mathcal{O}$ in the Hilbert space of the CFT, we need to specify its matrix elements in some reference basis. The natural basis, for dynamical reasons, is the energy basis $\ket{E_n}$ of the CFT Hamiltonian $H$. For the rest of the paper, we will work in this basis, and denote the matrix elements of the operator in this basis by $\mathcal{O}_{nm} =\bra{E_n} \mathcal{O} \ket{E_m}$.

Our main assumption is that of dynamical chaos in the high-energy spectrum of the CFT. This is encapsulated in the `ETH ansatz' for the matrix elements of the thin shell operator
\be\label{eq:randop}
\mathcal{O}_{nm}\, = \, e^{-S(\bar{E})/2}\,g(\bar{E},\omega)^{1/2}\,R_{nm}\;,
\ee
where, again, we have defined $\bar{E} = (E_n+E_m)/2$ and $\omega = E_n-E_m$. The smooth function $g(\bar{E},\omega)$ encodes the microcanonical two-point function of the operator, and the coefficients $R_{nm}$ are $O(1)$ complex numbers with erratically varying phases. In \eqref{eq:randop} we have neglected the smooth diagonal part, since the heavy shell operator lacks of an $O(1)$ microcanonical trace. 

%In the free bulk limit $G\rightarrow 0$, this can be seen from the fact that the operator $\mathcal{O}$ creates the dust particles in the bulk, while $\mathcal{O}^\dagger$ is responsible of annihilating them. In fact, since the operator $\mathcal{O}$ can only inject energy into the system, the envelope function will vanish for negative energy-differences, $g(\bar{E},\omega) \,\propto \;\Theta(\omega)$. 

The ansatz \eqref{eq:randop} requires further justification in this case, as the operator $\mathcal{O}$ is of system-size in the spatial sense, that is, it is completely delocalized over the $\mathbf{S}^{d-1}$. Moreover, the number of local operator insertions that constitute the operator scales with $N^2$ in the thermodynamic (large-$N$) limit. The argument which supports \eqref{eq:randop} is that the operator can be still regarded as simple enough in the internal large-$N$ space of the CFT to admit this form. Basically, its action consists in injecting a collection of particles in the asymptotic region of AdS. In the CFT, each of these particles is represented by an approximately local single-trace operator, so the shell is simply the product of local single-trace operators at different points along the sphere. This operator can still be considered relatively simple, as opposed to a much more generic polynomial of large multi-trace operators acting at each site.

However, exactly for the reason that the operator $\mathcal{O}$ that we consider is heavy, the matrix elements in \eqref{eq:randop} will display a different entropy-dependence to the one contained in the standard ETH form. Indeed, the envelope function $g(\bar{E},\omega)$ will now scale in some particular way with the microcanonical entropy $S(\bar{E})$, as well as with the rest mass $m$ of the thin shell. When inserted in thermal correlation functions, the envelope function $g(\bar{E},\omega)$ will modifying the large-$N$ saddlepoint equations that determine the dominant microcanonical contribution to the correlation function. As we will show, this phenomenon is dual to the gravitational backreaction exerted by the thin shell, that similarly modifies the semiclassical saddlepoint geometry contributing to the semiclassical correlation function.

In the regime in which $\bar{E}\gg  m$, the backreaction can be parametrically suppressed, namely by taking $m/\bar{E}\rightarrow 0$. In this regime the ansatz \eqref{eq:randop} acquires the standard ETH form. The simple character of the shell operator in this regime can be further diagnosed from holographic measures of operator size, such as suitable out-of-time-order correlators with probe operators \cite{Maldacena:2015waa}, or similarly in terms of notions of operator complexity, such as the Complexity=Volume proposal \cite{Stanford:2014jda}. These quantities display an initial `Lyapunov growth' under time-evolution for a thin shell perturbing a high-temperature thermal state. After some time, the operator becomes of `system-size', precisely at the scrambling time \cite{Shenker:2013pqa,Barbon:2019tuq, Susskind:2020gnl}.

	\subsection*{Boundary two-point function}

 In what follows, we will not provide any further justification to show that the ansatz \eqref{eq:randop} holds for the operator \eqref{eq:operator} at the level of its individual matrix elements. We will nevertheless be able to read off the typical magnitude of its matrix elements over the microcanonical band, captured in the smooth envelope function $g(\bar{E},\omega)$ up to $O(1)$ coefficients, by matching with the bulk thermal two-point function of the thin shell. To do that, let us first define
	\be\label{eq:Ffunction}
	f(E_n,E_m) = S(\bar{E}) - \log g(\bar{E},\omega)\,,
	\ee
	and rewrite the ansatz \eqref{eq:randop} in terms of this function
	\be\label{eq:randop2}
	\mathcal{O}_{nm}\, = \, e^{-f(E_n,E_m)/2}\,R_{nm}\;.
	\ee
	
	The object that we will analyze is the Euclidean two-point function \footnote{Upon analytic continuation $\tau \rightarrow it$, the Euclidean correlation function continues to the real time thermal correlation function $G_\beta(t)$. Moreover, the Euclidean correlation function also contains information about the two-sided correlation function $\bra{\text{TFD}}\mathcal{O}(t)_L\mathcal{O}^\dagger(0)_R\ket{\text{TFD}}$, obtained under a different analytic continuation $\tau \rightarrow \frac{\beta}{2}+it$.} 
	\be
	G_\beta(\tau) = \la \mathcal{O}(\tau) \mathcal{O}^\dagger(0)\ra_\beta =  \dfrac{1}{Z(\beta)}\sum_{n,m} e^{-(\beta-\tau) E_n} e^{-\tau E_m} |O_{nm}|^2\;.
	\ee
	
	Substituting the ansatz \eqref{eq:randop2} we can approximate the Euclidean two-point function replacing $|O_{nm}|^2$ by its typical value over the microcanonical window, given by the smooth envelope function $\exp (-f(E_n,E_m))$ up to $O(1)$ coefficients which we shall ignore. This yields
	\be\label{eq:twopointapprox1}
	G_\beta(\tau) \approx \dfrac{1}{Z(\beta)}\sum_{n,m} e^{-(\beta-\tau) E_n -\tau E_m - f(E_n,E_m)} \;.
	\ee
	
	The quantity \eqref{eq:twopointapprox1} now admits a smooth large-$N$ limit, a limit in which the level-spacing vanishes and we can replace the discrete sums in the previous expression by continuous integrals ($E_n  \rightarrow E$, $E_m \rightarrow E'$ ) using the continuous density of states $\rho(E) = E^{-1} \,e^{S(E)}$. The two-point function in this limit is given by the double integral
	\be\label{eq:twopointapprox2}
	G_\beta(\tau) \approx  \dfrac{1}{Z(\beta)}\int \dfrac{\text{d}E}{E} \int \dfrac{\text{d}E'}{E'} \,e^{S(E)+S(E')-(\beta-\tau)E-\tau E'-  f(E,E')}\;.
	\ee
	
	As we shall see, in order to consistently match the bulk physics, the integral must admit a microcanonical saddlepoint approximation as $N\rightarrow \infty$. The saddlepoint equations are simply 
	\begin{gather}
	\beta_E = \beta-\tau  \,+\,\partial_E f\,,\label{eq:saddle1}\\[.3cm]
	\beta_{E'} = \tau \,+\,\partial_{E'} f\;.\label{eq:saddle2}
	\end{gather}
	for $\beta_E = \frac{\text{d}S}{\text{d}E}$ the inverse temperature associated to the microcanonical band with energy $E$.
	
	%Compute specific heat
	
	Therefore, we find that at leading order in the asymptotic expansion $N\rightarrow \infty$, the the thermal two-point function is given by 
	\be\label{eq:classical2pt}
	G_\beta(\tau) \sim  e^{S(E)+S(E') - \beta F(\beta) -(\beta-\tau)E-\tau E'-  f(E,E')}\;
	\ee
	for $E,E'$ the microcanonical energies of the saddlepoint. The factor $F(\beta) =  -\beta^{-1}\log Z(\beta)$ corresponds to the thermodynamic free-energy, a factor which comes from the normalization of the two-point function. Here we are assuming that the saddlepoint is thermodynamically stable, of positive specific heat. Moreover, we will be solely focusing only on the `classical' value of the two-point function in the large-$N$ limit, ignoring the effect of one-loop determinants around these saddles.
	
	\subsection{Bulk two-point function}
	\label{sec:22}
	
	Using the standard form of the holographic dictionary, the two-point function $G_\beta(\tau)$ can likewise be evaluated semiclassically in the bulk. At leading order in the $G\rightarrow 0$ expansion, it will be given by 
	\be\label{eq:AdSCFTdict}
	G_\beta(\tau) \sim  \sum_{X}\,e^{-\Delta I[X]}\,
	\ee 
	where the sum is over manifolds $X$ respecting the asymptotic boundary conditions of the two-point correlation function at $\partial X$. Each saddle is weighted by $\Delta I[X] = I[X]-I[X_\beta]$, the renormalized action, where $I[X_\beta] = -\log Z(\beta)$ is the action of the Euclidean black hole $X_\beta$, which dominates the thermodynamics above the Hawking-Page temperature.
	
\begin{figure}[h]
		\centering
		\includegraphics[width=.385\textwidth]{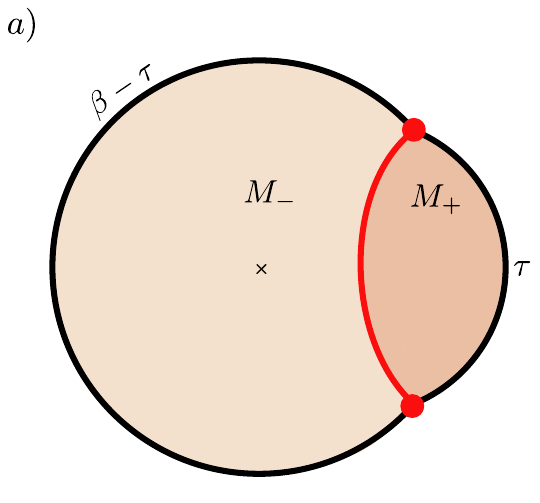}
		\hspace{.5cm}
		\includegraphics[width=.475\textwidth]{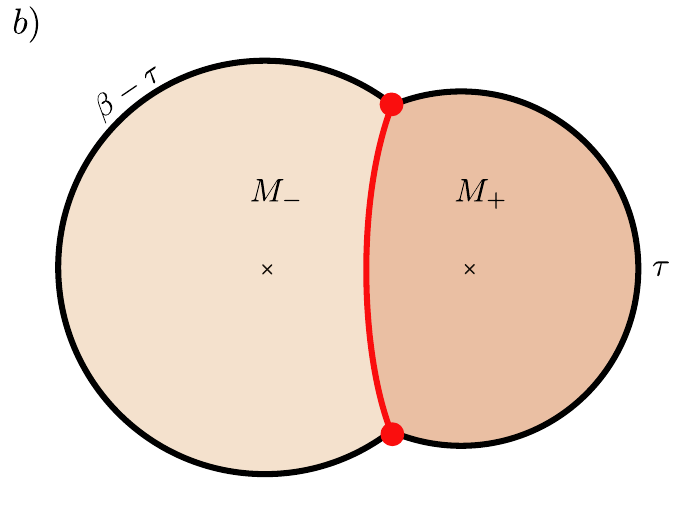}
		\caption{Geometry of the saddlepoint manifold $X$ for $\tau \in [0, \beta/2]$. The case $a)$ for $\tau <\tau_c$, in which the right patch $X^+$ does not include the tip of the right disk. The case $b)$ for $\tau >\tau_c$, in which the solution includes both tips.}
		\label{fig:bulk2pt}
	\end{figure}

	With our choice of operator $\mathcal{O}$, its effective bulk description is to insert a spherical thin shell of dust particles which propagates in the Euclidean section and backreacts on the geometry. At leading order in the $G\rightarrow 0$ expansion, the worldvolume $\mathscr{W}$ of the thin shell can only terminate at the insertion of $\mathcal{O}^\dagger$, which annihilates all the dust particles. Thence, the shell bisects the Euclidean manifold $X$ and generates two connected components $X^\pm \subset X$, one on each side of $\mathcal{W}$ (see Fig. \ref{fig:bulk2pt}).
	
	From spherical symmetry, the geometry of $X^\pm$ must be of the form
	\be\label{eq:geometry}
	\text{d}s_\pm^2\, = \,  f_\pm(r)\,\text{d}\tau_\pm \,+\,\dfrac{\text{d}r^2}{f_\pm(r)}\,+\,r^2\,\text{d}\Omega_{d-1}^2\;,
	\ee
	where, depending on the dimension, the blackening factor of each black hole is \footnote{We will conveniently set AdS units $\ell = 1$ throughout the rest of the paper.}
	\begin{gather}
	f_\pm(r)\, = \, r^2\,+\,1\,-\,\dfrac{16\pi G M_\pm }{(d-1) V_\Omega \,r^{d-2}}\,\hspace{.5cm}\text{for } d>2\label{metric}\\[.4cm]\label{blackeningtwo}
	f_\pm(r)\, = \, r^2\,-\,8GM_\pm\,\hspace{.5cm}\text{for } d=2\;.
	\end{gather}
	Here, $M_\pm$ is the ADM mass of the black hole, with inverse temperature $\beta_\pm = 4\pi/f'(r_\pm)$ and horizon radius $r_\pm$. The shell sits at $r=R(T)$, where $T$ is the synchronous proper time of the dust particles which form it.

	In the thin-shell formalism, the dynamics of the spherical thin shell gets reduced to the motion of a non-relativistic effective particle with zero total energy (see Appendix \ref{appendix:A})
	\be\label{eq:eomshell}
	\left(\dfrac{\text{d}R}{\text{d}T}\right)^2 + V_{\text{eff}}(R) = 0\;
	\ee
	subject to the effective potential
	\be
	V_{\text{eff}}(R) = -f_+(R) + \left(\dfrac{M_+-M_-}{m} - \dfrac{4\pi G m}{(d-1)V_\Omega R^{d-2}}\right)^2\;.
	\ee
	Qualitatively, the thin shell starts at $R = r_\infty$ and dives into the bulk, subject to a repulsive force in the Euclidean section. It bounces back at a minimum radius $R = R_* \geq r_\pm $ at the axis of time-reflection symmetry of the solution, and then gets back to $R = r_\infty$. The Euclidean time elapsed by the shell $\Delta \tau_\pm$ on each patch $X^\pm$ is \footnote{For details about the geometry and the thin shell formalism we refer the reader to Appendix \ref{appendix:A}.}
	\be\label{eq:shifttime}
	\Delta \tau_{\pm}  = 2 \int_{R_*}^{r_\infty}\dfrac{\text{d}R}{f_\pm(R)} \, \sqrt{\dfrac{f_\pm(R) + V_{\text{eff}}(R)}{- V_{\text{eff}}(R)}}\;.
	\ee
	
	Let us briefly comment on some of the general features of the saddlepoint manifold $X$ for $0\leq \tau \leq \frac{\beta}{2}$ (see Appendix \ref{appendix:A} for more details). For small values of $\tau$, the right patch $X^+$ will not include the tip of the Euclidean disk ($r=r_+$), and the geometry will look like case $a)$ of Fig. \ref{fig:bulk2pt}. As $\tau$ is increased, the shell explores more of the right cigar geometry and at some point, $\tau = \tau_c(\beta)$, the shell will intersect the tip of the right disk. For $\tau_c < \tau \leq \beta/2$, the solution $X$ corresponds to case $b)$ of Fig. \ref{fig:bulk2pt}. 
	
	These two different solutions $X$ admit a different interpretation in terms of the state which is prepared when the path integral is cut open along the axis of time-reflection symmetry. On the one hand, the situation for $\tau<\tau_c$ represents a two-sided state in which the shell sits outside the right horizon, while for $\tau > \tau_c$ the shell is trapped in the black hole interior. 
	
 For concreteness, let us here stick to the case $\tau > \tau_c$, and leave the analysis of the complementary regime for the discussion in Appendix \ref{appendix:A}. The gravitational `microcanonical saddlepoint' equations determine the pair of ADM energies $( M_-,M_+) $ in terms of the asymptotic thermal data $(\beta,\tau)$. They are simply given by the set of equations
	\begin{gather}
	\beta_- = \beta- \tau + \Delta \tau_-\;,\label{eq:bulksaddle1}\\[.2cm]
	\beta_+ =   \tau + \Delta \tau_+\,.\label{eq:bulksaddle2}\,
	\end{gather}
obtained from the identification of the periodicity in the coordinates $\tau_\pm \sim \tau_\pm + \beta_\pm$ of the solutions, in terms of the asymptotic data. For example, looking at Fig. \ref{fig:bulk2pt} $b)$,  the total Euclidean time periodicity of the right circle, $\beta_+$, corresponds to the boundary Euclidean time between the shell insertions, $\tau$, plus the Euclidean time elapsed by the shell on the right geometry, denoted by $\Delta \tau_+$.

	\subsubsection*{On-shell action}
	
\begin{figure}[h]
		\centering
		\includegraphics[width=.5\textwidth]{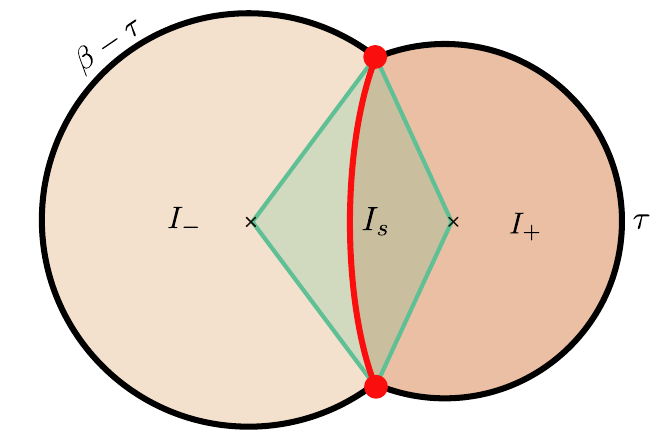}
		\caption{The different regions in the decomposition of the gravitational action $I[X]$. The green region $X_s$ accounts for the intrinsic contribution from the shell.}
		\label{fig:bulk2ptaction}
	\end{figure}

	In order to evaluate the bulk value of the two-point function in the classical limit \eqref{eq:AdSCFTdict}, we need to evaluate the renormalized action $I[X]$. To do that, it is convenient to divide the manifold into $X =X_- \cup X_s \cup X_+$ (see Fig. \ref{fig:bulk2ptaction}). From additivity, the on-shell action follows the decomposition 
	\be\label{eq:renormalbulkact}
	I[X] = I[X_-] + I[X_+] + I[X_s]\,.
	\ee
	
	The first two terms are intrinsic to the black holes and are given by 
	\be
	I[X_\pm]  = (\beta_\pm -\tau_\pm)F(\beta_\pm)\;,
	\ee
	where $F(\beta_\pm) = -\beta_\pm^{-1} \log Z(\beta_\pm)$ corresponds to the renormalized free energy of the respective black hole. After suitable holographic renormalization, it is given by (see e.g. \cite{Emparan:1999pm})
    \be
    F(\beta_\pm) = \dfrac{ V_\Omega}{8\pi G}\left(-r_\pm^{d}+ r_\pm^{d-2} + c_d\right)\;.
    \ee
    The constant $c_d$ accounts for the Casimir energy of the CFT in even dimensions \cite{Balasubramanian:1999re} ($c_d = -\frac{1}{2},\frac{3}{8}, -\frac{5}{16},\ldots$ in $d=2,4,6,\ldots$). 
	
	The last term $I[X_s]$, on the other hand, can be interpreted as the intrinsic contribution of the shell, since its value vanishes in the limit $m\rightarrow 0$. \footnote{In the $m\rightarrow 0$ limit with fixed $(\beta,\tau)$, the critical value $\tau_c\rightarrow \beta/2$ and this solution only exists at $\tau = \beta/2$ with $M_+ - M_- =0$, in which $I[X_s]\rightarrow 0$ clearly since $X_s$ shrinks to zero volume.} The Euclidean action associated to the region $X_s$ will have the form
	\be\label{eq:Ishell}
	I[X_s] = -\dfrac{1}{16\pi G}\int_{X_s} \,(R-2\Lambda)\,+\,\int_{\mathcal{W}} \sigma  \;,
	\ee
	before renormalization, where suitable counterterms need to be added to remove the long-distance divergences of both terms, for $r_\infty \rightarrow \infty$. The Einstein-Hilbert term produces two terms, since
	\be
	R-2\Lambda = -2d + \frac{16\pi G}{d-1}\delta(y)\,,
	\ee
	where $y$ is a normal coordinate to $\mathcal{W}$, sitting at $y=0$. Plugging this expression into \eqref{eq:Ishell} gives
	\be\label{eq:Ishell2}
	I[X_s] = \dfrac{d}{8\pi G}\,\text{Vol}(X_s)\,+\,m\dfrac{d-2}{d-1}  L[\gamma_{\mathcal{W}}]  \;,
	\ee	
	with $L[\gamma_{\mathcal{W}}]$ corresponding to the proper length of the trajectory of the heavy particle in the $(\tau_\pm,r)$ plane. Explicitly, each term is computed from the integrals
	\begin{gather}
	L[\gamma_{\mathcal{W}}] = 2\int_{R_*}^{r_\infty} \dfrac{\text{d}R}{\sqrt{-V_{\text{eff}}(R)}}\,,\label{eq:length}\\[.4cm]
	\text{Vol}(X^\pm _s) = \dfrac{2V_\Omega}{d} \int_{R_*}^{r_\infty}\dfrac{\text{d}R}{f_\pm(R)}\, \sqrt{\dfrac{f_\pm(R) + V_{\text{eff}}(R)}{- V_{\text{eff}}(R)}}\,(R^d-r_\pm^d)\;,\label{eq:vol}
	\end{gather}
	where $\text{Vol}(X_s) = \text{Vol}(X^+_s)+\text{Vol}(X^-_s)$. The solution for these integrals requires numerical treatment for $d>2$. 
	
	\subsubsection*{2+1 dimensions}
	
	For $d=2$, however, the second term drops out of \eqref{eq:Ishell2} and the volume of $X_s$ can be analytically computed, $\text{Vol}(X_s) = 4\pi G m L[\gamma_{\mathcal{W}}]$. This leads to the standard propagator of a massive particle
	\be\label{eq:2daction}
	I[X_s] = m L[\gamma_{\mathcal{W}}] = 2 m \,\text{cosh}^{-1}\left(\dfrac{r_\infty}{R_*}\right)\hspace{.8cm}\text{for } d=2 \,,
	\ee
	where we have evaluated the proper length for the explicit trajectory of the particle $R(T) = R_* \cosh T$, for $R_*^2 = r_+^2+ (\frac{M_+-M_-}{m}-2Gm)^2$. 
	
	To renormalize the logarithmic divergence of \eqref{eq:2daction} as $r_\infty \rightarrow \infty$, we choose to add the counterterm $I_{\text{ct}}[X_s] = -m \log r_\infty$ and then take $r_\infty \rightarrow \infty$. The renormalized action of the shell reads
	\be\label{eq:renshellact2d}
	I_{\text{ren}}[X_s]\,= \, - 2 m \log {R_*} + 2m \log 2 \hspace{.8cm}\text{for } d=2 \;.
	\ee
	
	\subsection{Completing the ansatz}
	
	Upon identifying $E = M_-$ and $E' = M_+$, the bulk saddlepoint equations \eqref{eq:bulksaddle1} and \eqref{eq:bulksaddle2} structurally reduce to the microcanonical saddlepoint equations \eqref{eq:saddle1} and \eqref{eq:saddle2} in the CFT. \footnote{We have only focused on the high-temperature two-point function, which is dominated by $X$ constructed from large AdS black hole solutions. This will be enough to get the envelope function of the thin shell in the high-energy spectrum.}  Comparing them allows to identify the value of the envelope function $f(E,E')$ for the shell operator, which must satisfy 
	\begin{equation}\label{eq:cc}
		\Delta \tau_\pm = \partial_\pm f(E,E')\;.
	\end{equation}
	We note that it is therefore non-trivial that such a function $f(E,E')$ exists at all, and to a large extent \eqref{eq:cc} is a test of the consistency between the approximation \eqref{eq:twopointapprox1} and the gravitational calculation. Indeed, for a smooth solution of \eqref{eq:cc} to exist, its mixed second derivatives must coincide
	\be
	\partial_- \Delta \tau_+ = \partial_+ \Delta \tau_-\;.
	\ee
	Using the expressions \eqref{eq:shifttime} directly, it is straightforward to verify this relation does in fact hold in any dimension. We thus find that the gravitational calculation of the two-point function is consistent with the form of \eqref{eq:randop2} for the magnitude of typical matrix elements in the microcanonical band, at leading order in the large-$N$ expansion.
	
	In order to find the explicit form of $f(E,E')$, we choose to proceed by comparing the microcanonical saddlepoint action in \eqref{eq:classical2pt} to the renormalized gravitational action \eqref{eq:renormalbulkact}, with the suitable subtraction of $I[X_\beta]$ associated to the normalization of the two-point function. The resulting envelope function is given by \footnote{The derivation of $f(E,E')$ presented in this section is strictly valid for the regime of small energy differences $\omega < \omega_c(\bar{E})$ (see Appendix \ref{appendix:A}). The same form of $f(E,E')$ also extends for the complementary regime $\omega \geq \omega_c(\bar{E})$, as long as the microcanonical bands associated to $\bar{E}$ is dominated by large black hole microstates.}
	\be\label{eq:functionF}
	f(E,E') \, = \,\alpha_- S(E)\,+\,\alpha_+ S(E')\,+\,I_s(E,E')\;,
	\ee
	for $I_s(E,E') = I[X_s]$ in \eqref{eq:Ishell2} and for the ${O}(1)$ coefficients $\alpha_\pm = \frac{\Delta\tau_\pm}{\beta_\pm}$. The value of these coefficients is energy-dependent, and it ranges from $0\leq \alpha_\pm \leq \frac{1}{2}$, where the lower limit corresponds to small energies $\bar{E} \ll m $, while the upper limit requires large energies $\bar{E} \gg m$, relative to the rest mass of the shell.

\begin{figure}[h]
		\centering
		\includegraphics[width=.8\textwidth]{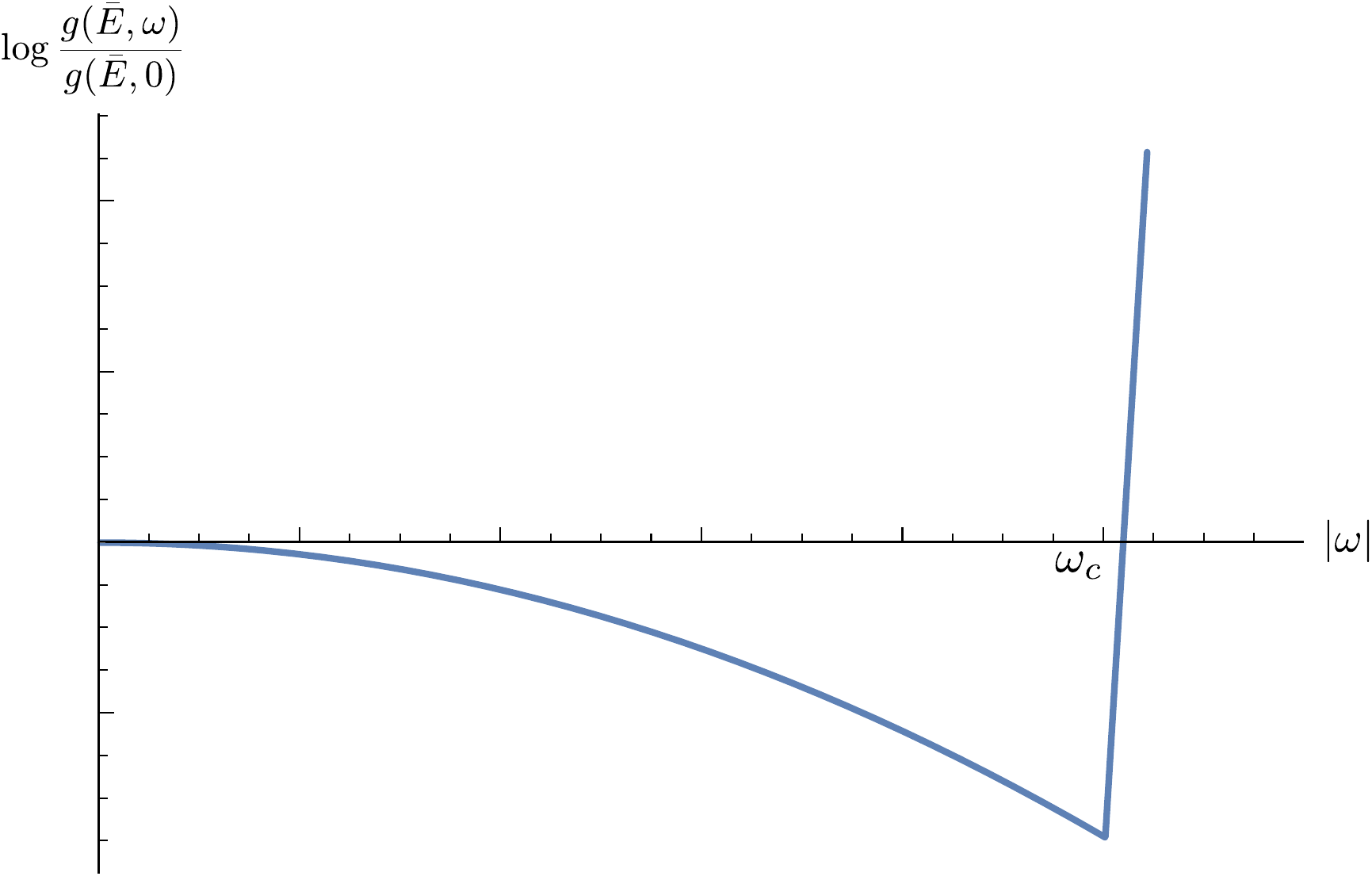}
		\caption{The numerical value of the envelope function as a function of the energy difference $|\omega|$ for the thin shell operator in $d=2$. In this case the critical value is $\omega_c = Gm^2 \sim m$, independent of $\bar{E}$. }
		\label{fig:g}
	\end{figure}

	In terms of the original envelope function $g(\bar{E},\omega)$ present in the form of the generalized ETH ansatz \eqref{eq:randop}, our result translates to
	\be\label{eq:gsol}
	\log g(\bar{E},\omega) \, = \, S(\bar{E}) - \alpha_- S(\bar{E}-\omega) - \alpha_+ S(\bar{E}+\omega) -I_s(\bar{E},\omega)\;.
	\ee
	The value of the envelope function of the thin shell is plotted in Fig. \ref{fig:g}. The function will be extremely peaked at an energy-difference $\omega = \omega_\infty \rightarrow \infty$ associated with the total energy of of inserting a collection of particles arbitrarily close to the asymptotic boundary of AdS. The operator has been renormalized by the addition of suitable counterterms to the gravitational action, in such a way that all the UV divergences in the $r_\infty \rightarrow \infty$ limit are regulated. The tail in energy differences provides the information about the variance in energy of the thin shell operator, which creates a semiclassical bulk state. There is a second much smaller peak at $\omega =0$, and a change in tendency at $|\omega| = \omega_c \sim m$ associated to the transition between both gravitational saddles. The envelope function is an even function of $\omega$. 
	
	We note that for high-energies compared to the mass of the shell, $\bar{E}\gg m$, the envelope function \eqref{eq:gsol} is approximated by $g(\bar{E},\omega) \approx \exp ( -I_s(E,E'))$, which is the part of the gravitational action intrinsic to the thin shell, proportional to its rest mass. Therefore, in this regime, the matrix elements of the shell in the ansatz \eqref{eq:randop} do strictly follow the entropy-suppression of the standard ETH ansatz for a simple operator. At intermediate energy regimes $\bar{E}\lesssim m$, on the other hand, the function $g(\bar{E},\omega)$ does in fact modify the entropy-dependence of the matrix elements in the ansatz \eqref{eq:randop}.

	\section{Wormholes from heavy operator statistics}
	\label{sec:3}
	
	The analysis of the previous section explicitly shows that the typical magnitude of the microcanonical matrix element of the shell operator, encapsulated in the envelope function $g(\bar{E},\omega)$ of the `ETH ansatz' \eqref{eq:randop}, is consistent with the physics of the bulk thermal two-point function at high-temperatures. In particular, the operator is able to modify the structure of the microcanonical saddlepoint dominating the thermal correlation function. In the bulk, this process is dual to the classical backreaction of the thin shell, that modifies the Euclidean saddlepoint geometry.
	
	We remark that the semiclassical value of the correlator is agnostic about the fine-grained details of exact two-point function $G_\beta(\tau)$, which can even be sensitive to individual matrix elements of the operator. Instead, the semiclassical path integral effectively computes the coarse-grained correlation function \eqref{eq:twopointapprox1}, where matrix element $| \mathcal{O}_{nm}|^2$ has been replaced by its typical value over the microcanonical window, in this case controlled by the envelope function in the ansatz \eqref{eq:randop2}. Furthermore, the smooth semiclassical bulk description requires to take $N\rightarrow \infty$, a limit in which the spectral level-spacing vanishes and the sums over energy levels can be replaced by integrals over continuous functions. 
	
	A more radical coarse-graining of the operator can be implemented by starting from the ETH ansatz \eqref{eq:randop} for its matrix elements. The ansatz assumes that all of its individual matrix elements democratically encode its microcanonical expectation value, with the addition of very erratic $O(1)$ complex numbers $R_{nm}$ that depend in the microscopic description of the operator. From this point of view, we can \textit{define} an ensemble of operators by promoting the erratic coefficients $R_{nm}$ in \eqref{eq:randop} to independent complex random variables of zero mean an unit variance. Provided that $g(\bar{E},\omega)$ is given by \eqref{eq:gsol}, the members of this ensemble of operators share the thermal correlation functions with the original thin shell operator and are thus semiclassically indistinguishable, at least within the low-energy effective description in terms of Einstein gravity coupled to the thin shell.

	In this section, we show that semiclassical gravity provides an effective statistical description of the thin shell operator in terms of the ETH ensemble of operators that we just defined. In particular, we analyze the cumulants of the thermal correlation functions over the ensemble of operators. Since we have determined the value of the envelope function $g(\bar{E},\omega)$ in the ensemble, we will be able to evaluate the cumulants in a microcanonical saddlepoint approximation in the large-$N$ limit. We then proceed to construct Euclidean wormhole solutions that provide connected contributions to products of correlation functions, and that should be included according to the rules of the semiclassical path integral of gravity. We finally show that the sadddlepoint equations of these wormholes exactly match the microcanonical saddlepoint equations for the cumulants. Similarly, the value of the gravitational action of these wormholes matches the answer for the cumulants in the large-$N$ limit.

	\subsection{One-point function}
	\label{sec:31}
	
	Let us start by considering thermal one-point function $ \la \mathcal{O}\ra_{\beta} = Z(\beta)^{-1}\text{Tr}(e^{-\beta H} \mathcal{O})$. In components it reads
	\be\label{eq:oneptfn}
	\la \mathcal{O}\ra_{\beta} = \dfrac{1}{Z(\beta)}\sum_{n} e^{-\beta E_n} \mathcal{O}_{nn}\,.
	\ee
	Using the ansatz \eqref{eq:randop2} it is clear that the value of the one-point function is highly sensitive to the fine-grained phases $R_{nm}$. It averages to zero within the ensemble of operators, $\overline{\la \mathcal{O}\ra_{\beta}} = 0$. This agrees with the output of the semiclassical path integral, in which the shell lacks of a bulk one-point function in the $G\rightarrow 0$ limit.
	
	Instead, we shall now consider the product of one-point functions
	\be
	\la \mathcal{O}\ra_{\beta}\la \mathcal{O}^\dagger \ra_{\beta'} = \dfrac{1}{Z(\beta)Z(\beta')}\sum_{n,m} e^{-\beta E_n-\beta E_m} \mathcal{O}_{nn}\mathcal{O}_{mm}^*\,,
	\ee
	 taken in all generality at different temperatures. Plugging in the ETH form \eqref{eq:randop2}, this quantity can now be approximated by a `phase-correlated' amplitude, arising from the statistical variance of the random coefficients $R_{nm}$, which we assume are independent, $ R_{nn}R_{mm}^* \approx \overline{R_{nn}R_{mm}^*} = \delta_{nm}$. The variance over the ensemble of operators is given by 
	\be\label{eq:varianceonepointbefore}
	 \overline{\la \mathcal{O}\ra_{\beta}\la \mathcal{O}^\dagger \ra_{\beta'}} =  \dfrac{1}{Z(\beta)Z(\beta')}\sum_{n} e^{-(\beta + \beta') E_n - f(E_n,E_n) }\;.
	\ee

    Note that in this case the observable \eqref{eq:oneptfn} only involves the diagonal matrix elements of the operator in the energy basis. Therefore, we can set $\omega = 0$ in the envelope function $g(\bar{E},\omega)$ to compute the value of this observable and its higher moments over the ensemble. This is the reason why there is only the diagonal part of $f(E_n,E_m)$ appearing in the expression \eqref{eq:varianceonepointbefore}. More general observables which we will discuss later, such as thermal two-point function $G_\beta(\tau)$, require of the complete envelope function $g(\bar{E},\omega)$, or equivalently, of the function $f(E_n,E_m)$.

	In the large-$N$ limit, \eqref{eq:varianceonepointbefore} admits the integral representation
	\be
	\overline{\la \mathcal{O}\ra_{\beta}\la \mathcal{O}^\dagger \ra_{\beta'}}  \sim  \dfrac{1}{Z(\beta)Z(\beta')} \int \dfrac{\text{d}E}{E}\,e^{-(\beta + \beta')E + S(E) - f(E,E)}\;.
	\ee

	Since the envelope function $f(E,E')$ has been determined, and is given by \eqref{eq:functionF}, we can now show that this integral admits a microcanonical saddlepoint approximation in the large-$N$ limit of the CFT. Using the expression \eqref{eq:cc}, we learn that the dominant microcanonical window must be centered at an energy $E$ which solves
	\be\label{eq:saddlewh}
	\beta_E = \beta + \beta' + 2 \Delta \tau\;,
	\ee
	where we have defined $\Delta \tau = \Delta \tau_+= \Delta \tau_+$ from \eqref{eq:shifttime} with $M_+ = M_- = E$.

	In this way, the product of one-point functions is given by its microcanonical saddlepoint value at leading order in the large-$N$ expansion 
	\be\label{eq:bulk1ptwh}
	\overline{\la \mathcal{O}\ra_{\beta}\la \mathcal{O}^\dagger \ra_{\beta'}}  \sim e^{-(\beta + \beta')E + S(E) - f(E,E) + \beta F(\beta) +  \beta' F(\beta') }\;,
	\ee
	where again $\beta F(\beta)  = -\log Z(\beta)$ is the thermodynamic free-energy of the CFT, coming from the normalization of the one-point function.

     \subsubsection*{Euclidean wormhole solution}
	
	We remark that the expression \eqref{eq:bulk1ptwh} is predicting a microcanonical saddlepoint contribution for the variance of the one-point function over the microscopic ensemble of ETH operators. We will now explicitly reproduce such a quantity from the gravitational computation of the square of the one-point function, sticking to the rules of the semiclassical path integral of gravity \eqref{eq:semiclassicalpi}. In particular, we will show that the variance is captured in the $G\rightarrow 0$ expansion by an Euclidean wormhole $X$ contributing to the square of the one-point function
	\be\label{eq:whstatistics}
	\overline{\la \mathcal{O}\ra_{\beta}\la \mathcal{O}^\dagger \ra_{\beta'}} \sim  e^{-\Delta I[X]}\,,
	\ee
	where $\Delta I[X]$ is the renormalized gravitational action of the wormhole.

	According to the rules of the semiclassical path integral, the asymptotic boundary conditions that prepare the product $\la \mathcal{O}\ra_{\beta}\la \mathcal{O}^\dagger \ra_{\beta'}$ correspond to two asymptotic thermal circles, $M_1 = \mathbf{S}_{\beta}^1 \times \mathbf{S}^{d-1}$ and $M_2= \mathbf{S}_{\beta'}^1 \times \mathbf{S}^{d-1}$, with the respective operator insertion of $\mathcal{O}$ or $\mathcal{O}^\dagger$ at each boundary.  We now need to fill the bulk geometry in with all of the allowable solutions given these boundary conditions. The connected contribution vanishes at leading order, from the fact that the thin shell lacks of a thermal one-point function at this order.
	
	The wormhole $X$ that we are seeking must be of topology $\mathbf{R} \times \mathbf{S}^1 \times  \mathbf{S}^{d-1}$ with $\partial X = M_1 \cup M_2$, the disjoint union of the two boundaries. It is well-known that such solutions do not exist in pure Einstein gravity \cite{Witten:1999xp}. Here, the effect of the heavy spherical shell is crucial to stabilize the wormhole.  The shell propagates between the operator insertions on both boundaries, and the localized backreaction exerted on its worldvolume $\mathcal{W}$ is responsible of stabilizing the wormhole and creating a valid solution $X$ of the gravitational field equations. \footnote{Heavy particles have been previously considered as a stabilization mechanism for wormholes in studies of semiclassical wormhole solutions in JT gravity \cite{Stanford,Hsin:2020mfa}.}

	 From spherical symmetry it is clear that the geometry of the wormhole $X$ must be locally of AdS-Schwarzschild type away from the shell
	 \be\label{eq:metricwh}
	 \text{d}s_X^2\, = \, f(r) \text{d}\tau^2 + \dfrac{\text{d}r^2}{f(r)} + r^2\text{d}\Omega_{d-1}^2\;,
	 \ee
    where, depending on the dimension, $f(r)$ corresponds to the blackening factor \eqref{metric} or \eqref{blackeningtwo}.
     
\begin{figure}[h]
		\centering
		\includegraphics[width=.33\textwidth]{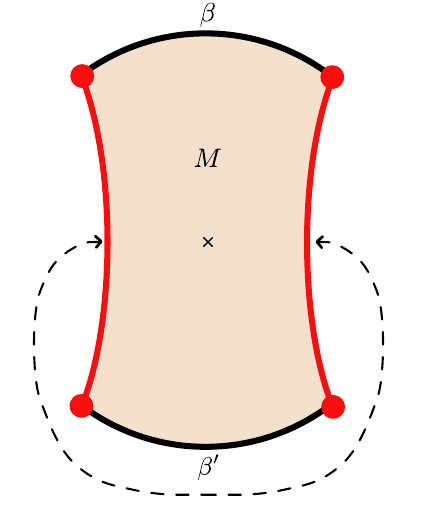}
		\hspace{2cm}
		\includegraphics[width=.21\textwidth]{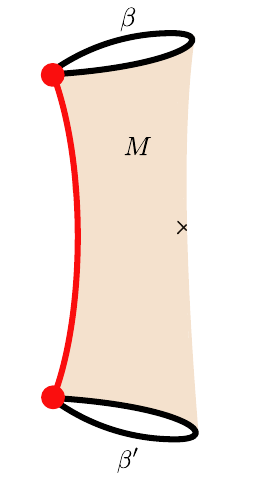}
		\caption{The wormhole $X$ is deconstructed on the left figure. It corresponds to a portion of an Euclidean black hole, cut by the trajectories of two thin shells, which are then identified. On the right, the resulting $\mathbf{R}\times  \mathbf{S}^1\times \mathbf{S}^{d-1}$ topology connecting both boundaries. The mass $M$ of the wormhole only depends on the combination $\beta+\beta'$.}
		\label{fig:1ptwh}
	\end{figure}

	 The way to construct the solution $X$ is to first use two copies of $\mathcal{W}$ to cut the Euclidean black hole along the trajectories of two thin shells. The second step is to glue the two copies of $\mathcal{W}$ and wrap the Euclidean disk to form the transverse $\mathbf{S}^1$ of the wormhole $X$ (see Fig. \ref{fig:1ptwh}). The processes of cutting and gluing can be analyzed within the general thin shell formalism reviewed in Appendix \ref{appendix:A}, taking the same mass for the two patches, $M_\pm = M$. 
	 
	The effective potential of the shell in this case is
	\be
	V_{\text{eff}}(R) = -f(R) + \left(\dfrac{4Gm}{(d-1)V_\Omega R^{d-2}}\right)^2\,,
	\ee
	Hence, $R_* > r_+$, and the wormhole has a non-vanishing section only as a consequence of the gravitational self-energy of the shell. The minimal proper section of the wormhole will have volume
	\be
	\Delta V \approx 2\int_{r_+}^{R_*} \dfrac{R^{d-1}dR}{\sqrt{f(R)}} \approx \dfrac{2Gm \beta_M r_+ }{(d-1)\pi V_\Omega }\:,
	\ee 
    and this volume will be parametrically large in Planck units provided that the shell is heavy, $m \sim G^{-1}$.

	The microcanonical equation that determines the mass $M$ of the wormhole can be directly extracted from the geometry in Fig. \ref{fig:1ptwh}. The Euclidean periodicity of the solution is determined to be
	\be\label{eq:spewh}
	\beta_M = \beta + \beta' + 2\Delta \tau\;,
	\ee
	where $\beta + \beta'$ is the part coming from the two asymptotic boundaries, while $2\Delta \tau$ corresponds to the part of the two cuts of the Euclidean disk. The mass $M$ will only depend in the combination $\beta + \beta'$. 
	
	Therefore, comparing this equation with \eqref{eq:saddlewh}, we observe that the microcanonical saddlepoint equation of the wormhole coincides with the microcanonical saddlepoint equation for the variance of the one-point function over the microscopic ensemble of operators in the CFT. 
 
    The renormalized action of the wormhole $I[X]$ can be evaluated in the same way as in the previous section, where it was found convenient to divide $X$ into different regions to isolate the intrinsic contribution from the shell. In this case the total action is
	\be\label{eq:actionwh}
	I[X] = (\beta + \beta') F(\beta_M) \,+ I[X_s]\;,
	\ee
	where $F(\beta_M) = -\beta_M^{-1}\log Z(\beta_M)$ is the free energy of the Euclidean black hole that we start from, and $I[X_s]$ is given by \eqref{eq:Ishell2} with $M_\pm = M$.
	
	Adding the normalization factors, $\Delta I[X] =  I[X] - I[X_\beta] - I[X_{\beta'}]$, the classical contribution of the wormhole
	\be\label{eq:bulk1ptwhbulk}
	e^{-\Delta I[X]} = e^{-(\beta + \beta') F(\beta_M)\,+ I[X_s] + \beta F(\beta) +  \beta' F(\beta') }\;,
	\ee
	matches the statistical average of the product of one-point functions in the CFT \eqref{eq:bulk1ptwh}, which can be checked from the particular form of the envelope function \eqref{eq:functionF}. We conclude that the wormhole $X$ successfully accounts for the connected contribution in the statistical description of the product of one-point functions:
	\be\label{eq:whstatistics2}
	\overline{\la \mathcal{O}\ra_{\beta}\la \mathcal{O}^\dagger \ra_{\beta'}} \,\sim\, e^{-\Delta I[X]}\,.
	\ee
    This shows that the output of the semiclassical path integral for the product of one-point functions of the thin shell is a statistical characterization of the real product of one-point functions, averaged over an ensemble of microscopic operators compatible with the effective bulk description of the thin shell.

	\subsubsection*{2+1 dimensions}
	
	To gain some analytic intuition about the wormhole contribution $\Delta I[X]$, let us evaluate it in $d=2$ spatial dimensions, at equal temperatures $\beta = \beta'$ for concreteness. The free-energy of the BTZ black hole is $F(\beta_M) = -\frac{\pi^ 2}{2G\beta_M^ 2}$, while the renormalized action of the shell is given by \eqref{eq:renshellact2d}. The total action is then
	\be\label{eq:actionwh2d}
	I[X] = -\dfrac{\pi^ 2\beta}{G\beta_M^ 2} \,- 2m \log \dfrac{R_*}{2} \;,
	\ee
	where $R_* = \sqrt{r_+^ 2+ (2Gm)^ 2}$.

	In the high-temperature regime $\beta  \ll (Gm)^{-1}$, we can approximate $R_* \approx r_+ = \frac{2\pi}{\beta_{M}}$ and the euclidean time elapsed by the shell is $2\Delta \tau \approx \beta_{M} - \beta_M^2 \frac{2Gm}{\pi^2}$ according to the expansion of \eqref{eq:app2+1sol}. The micocanonical equation \eqref{eq:spewh} can now be solved for the inverse temperature of the wormhole
	\be
	\beta_M \approx \sqrt{\dfrac{\pi^2\beta }{Gm}}\;.
	\ee
	Plugging the temperature back in \eqref{eq:actionwh2d} we get that the proper wormhole contribution is constant in the $\beta \rightarrow 0$ limit, and that the total action is dominated by the intrinsic action of the shell. Adding the normalization this gives
	\be
	\Delta I[X] \sim   4m \log \beta - \dfrac{\pi^ 2}{G\beta}\hspace{.5cm}\text{as }\beta\rightarrow 0\;.
	\ee
	So, even if the  wormhole amplitude, given by $\exp(-I[X])$, becomes large at high energies, the total contribution to \eqref{eq:wormholehightemp} is suppressed by the normalization, and it scales exactly as the Cardy entropy
	\be\label{eq:wormholehightemp}
	\overline{\la \mathcal{O}\ra_{\beta}\la \mathcal{O}^\dagger \ra_{\beta}} \sim e^{-S(\beta)}\hspace{.5cm}\text{as }\beta\rightarrow 0\;.
	\ee

\begin{figure}[h]
		\centering
		\includegraphics[width=.7\textwidth]{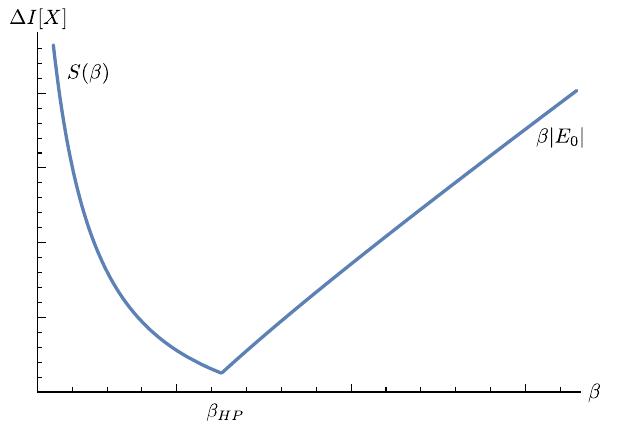}
		\caption{Renormalized action of the wormhole $\Delta I[X]$ as a function of the temperature, for $d=2$. The Hawking-Page transition at $\beta_{HP}=2\pi$ affects the normalization of the amplitude.}
		\label{fig:whaction}
	\end{figure}

	On the other hand, at low-temperatures $\beta \gg 2\pi $, the microcanonical solution is $\beta_M \approx 2\beta $ and the action \eqref{eq:actionwh2d} becomes finite as $\beta\rightarrow \infty$. The normalization however is dominated by the Casimir energy, giving a total contribution
	\be\label{eq:wormholehightemp2}
	\overline{\la \mathcal{O}\ra_{\beta}\la \mathcal{O}^\dagger \ra_{\beta}} \sim  e^{2\beta E_0}\hspace{.5cm}\text{as }\beta\rightarrow \infty\;,
	\ee
	where $E_0 = -(8G)^{-1}$. Note that the normalization here comes from the dominant AdS saddlepoint. Thus the connected contribution is also largely suppressed at low temperatures. There is an intermediate regime around the Hawing-Page temperature $\beta = 2\pi$ in which the wormhole becomes less suppressed (see Fig. \ref{fig:whaction}).
	
	\subsubsection*{Comment on analytic continuation}
	
	For different boundary moduli $\beta \neq \beta'$, the wormhole $X$ possesses a single axis of reflection-symmetry (vertical axis in Fig. \ref{fig:1ptwh}). The Euclidean solution analytically continues to the two-sided Schwarzschild-AdS black hole, and the trajectory of the shell becomes complex under this analytic continuation. 
	
	For $\beta = \beta'$, there is a second axis of reflection-symmetry (horizontal axis in Fig. \ref{fig:1ptwh}), under which the boundaries are mapped to each other. Along this axis, the wormhole is analytically continued to a closed `AdS cosmology' (similar to those in \cite{Antonini:2022xzo,Antonini:2022opp}). The cosmology is constructed by cutting and gluing the two-sided Schwarzschild-AdS solution along the the trajectories of two thin shells which leave the white hole region to the left/right exterior regions respectively, and then re-enter the black hole.
	
	\subsection{A Hawking-Page wormhole transition}
	\label{sec:32}
	
	The `black wormhole' constructed above, with $M> 0$, is a thermodynamically stable solution as long as the AdS black hole that it is constructed from is large, of positive specific heat $C_V^{-1}  =  - \beta^{-2}\partial_E^2S >0$. This property follows from the convexity of the envelope function $\partial_E^2 f(E,E) = \partial_E \Delta \tau > 0$, which can be checked directly from the general expression \eqref{eq:shifttime}.  In this sense, it is a legitimate saddle contributing to the semiclassical amplitude. From this point on, we shall refer to this solution as $X_{\text{bh}}$ for the prominent role played by the Euclidean black hole.
	
	From standard considerations of thermodynamics in AdS space, we might expect that this wormhole will not provide the dominant contribution at low temperatures, somewhat below the analog of the Hawking-Page temperature for the wormhole. The question is whether there will exist a `vacuum AdS' wormhole solution $X_{\text{AdS}}$ with smaller renormalized action $I[X_{\text{AdS}}]<I[X_{\text{BH}}]$ in this regime.
	
\begin{figure}[h]
		\centering
		\includegraphics[width=.3\textwidth]{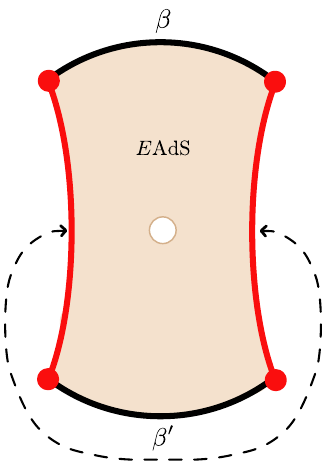}
		\hspace{2.5cm}
		\includegraphics[width=.18\textwidth]{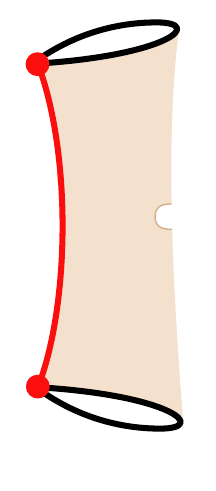}
		\caption{Deconstruction of the AdS wormhole. The geometry corresponds to a portion of Euclidean AdS, cut by the trajectories of two thin shells, which are then identified. The periodicity $\beta_{\text{AdS}}$ of the AdS solution only depends on the combination $\beta+\beta'$.}
		\label{fig:1adswh}
	\end{figure}

	The answer can be shown to be affirmative in a rather simple way. The (de)construction of the AdS wormhole $X_{\text{AdS}}$ is very similar to the one for $X_{\text{bh}}$. It begins by considering the $M=0$ limit of \eqref{eq:metricwh}. Topologically, the thermal $\mathbf{S}^1$ is now non-contractible in the bulk, for the reason that spatial spheres now shrink to zero size at $r=0$. Moreover the value of the Euclidean periodicity $\beta_{\text{AdS}}$ now becomes independent of the classical ADM mass of the solution, since the back-reaction of the thermal gas of particles in AdS can be neglected to leading order. The process to build the wormhole is again to cut two patches of the Euclidean space, along the trajectories of two spherical thin shells, and then to glue them together to form the connected manifold with disconnected boundaries (see Fig. \ref{fig:1adswh}). 
	
	In this case, the equation of motion of the shell \eqref{eq:eomshell} is governed by the effective potential
	\be
	V_{\text{eff}}(R) = -1 -R^2 + \left(\dfrac{4Gm}{(d-1)V_\Omega R^{d-2}}\right)^2\,.
	\ee
	The shell bounces back inside its Schwarzschild radius, at $R_*>0$, due to its gravitational self-energy. The solution can be trusted since the mass density $\sigma =\frac{m}{V_\Omega R^{d-1}}$ remains below the Planck density at $R=R_*$, for shells with mass $m$ parametrically larger than the Planck mass. \footnote{ In $d=2$ the solution only bounces back at a finite radius for heavy shells satisfying $2Gm \geq 1$.}
		
	The microcanonical saddlepoint equations for this wormhole take the form
	\be
	\beta_{\text{AdS}} = \beta +\beta' + 2\Delta\tau_0
	\ee
	where $\Delta \tau_0 = \Delta \tau_{\pm}$ is now a constant, independent of $\beta+\beta'$, obtained from \eqref{eq:shifttime} with $M_\pm =0$.
	
	The classical action of the AdS wormhole is likewise
	\be
	I[X_{\text{AdS}}] =   (\beta + \beta') F(\beta_{\text{AdS}})\,+\,I[X_s]\;,
	\ee
	where $F(\beta_{\text{AdS}}) = \beta_{\text{AdS}}^{-1} I_{\text{AdS}}$ is the renormalized free-energy of thermal AdS (cf. \cite{Emparan:1999pm}). The value of the shell's action, $I[X_s]$ is a constant, independent of $\beta+\beta'$, given by \eqref{eq:Ishell2}.
	
	We thus find a second contribution to the semiclassical product of correlation functions, which acquires the form 
	\be\label{eq:twowh}
	\overline{\la \mathcal{O}\ra_{\beta}\la \mathcal{O}^\dagger \ra_{\beta'}}  \sim e^{-\Delta I[X_{\text{bh}}]} + e^{-\Delta I[X_{\text{AdS}}]}\;.
	\ee
	
\begin{figure}[h]
		\centering
		\includegraphics[width=.95\textwidth]{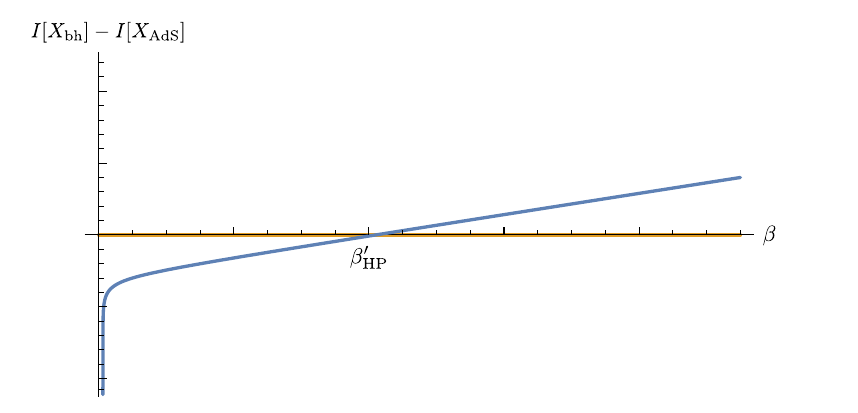}
		\caption{There is an exchange in dominance between the black wormhole $X_{\text{bh}}$ and the AdS wormhole $X_{\text{AdS}}$, at $\beta = \beta'_\text{HP}$. In $d=2$, we have numerically checked that $\beta'_\text{HP}\geq 2\pi$, and  $\text{d}\beta'_\text{HP}/\text{d}m >0$.}
		\label{fig:hptrans}
	\end{figure}

	Numerically we have checked that there is a `Hawking-Page transition' for the wormhole (see Fig. \ref{fig:hptrans}). The effect of the heavy shell is to slightly modify the details of the transition. In particular, for $d=2$, we numerically observe that the wormhole transition temperature is below the actual Hawking-Page temperature $\beta_{\text{HP}}' \geq 2\pi $.
	
	\subsubsection*{Comment on analytic continuation}
	
	For different boundary moduli $\beta \neq \beta'$, the wormhole $X_{\text{AdS}}$ possesses a single axis of reflection-symmetry (vertical axis in Fig. \ref{fig:1adswh}). The Euclidean solution analytically continues to two disconnected thermal AdS spacetimes, and the trajectory of the shell becomes complex under this analytic continuation. 
	
	For $\beta = \beta'$, however, there is a second axis of reflection-symmetry (horizontal axis in Fig. \ref{fig:1adswh}), under which the boundaries are mapped to each other. Along this axis, the wormhole is analytically continued to a second closed `AdS cosmology'. The cosmology is constructed by cutting and gluing two copies of vacuum AdS along the the trajectories of two thin shells, which contract towards the past/future.
	
	\subsubsection*{Averaging below Hawking-Page}	
	
	The semiclassical wormhole contribution from $X_{\text{AdS}}$ can be analogously reproduced in terms of averaging over an ensemble of microscopic operators, under the assumption that the ETH form \eqref{eq:randop} for the matrix elements of the heavy shell extends into the low-lying sector of the holographic system. It is true that the thermal gas of particles becomes integrable in strict large-$N$ limit of the CFT, and as operator $\phi$ which creates a single dust particle will definitely become sparse in the energy basis. Such an operator is expected to develop an ETH form at subleading orders in $1/N$, when the bulk interactions are turned on. The shell operator, on the other hand, is composed of a number of particles which scales with $N^2$, and as such, we expect that it respects the `ETH form' even at leading order. Moreover, as we showed in section \ref{sec:2}, the envelope function of the shell modifies the entropy-dependence in the ETH ansatz at low energies compared to $m$, so we expect that the low-lying matrix elements are still suppressed by $e^{-m} \sim e^{-N^ 2}$.
	
	These results suggests that there is also averaging below the Hawking-Page temperature for the shell operator, in fact dominated by the low-lying microcanonical bands in AdS. We should however not interpret these results as a counter-example of \cite{Schlenker:2022dyo}. The reason is that we have inserted a heavy operator whose energy scales with $N^2$, and thus its matrix elements cannot be strictly considered light sub-threshold observables in the large-$N$ limit. This is similar, but perhaps more manifest in this case, to the situation for the operators considered in \cite{Chandra:2022bqq} which create heavy point-particles in AdS$_3$.

	\subsection{Higher-point functions and multi-boundary wormholes}
	
	The construction of the wormhole presented above can be trivially extended to higher-point functions of the shell operator. Consider for concreteness the product of thermal two-point functions
	\be\label{eq:product2pt}
	G_\beta(\tau)\, G_{\beta'}(\tau') \,  = \, \dfrac{1}{Z(\beta)Z(\beta')} \sum_{n,m} e^{-(\beta-\tau) E_n - \tau E_m -(\beta'-\tau') E_p - \tau E_q}  \,|\mathcal{O}_{nm}|^2|\mathcal{O}_{pq}|^2\;.
	\ee
	
	We will assume that the ETH ensemble of operators is characterized by Gaussian statistics for the $R_{nm}$ coefficients, at leading order in the $N\rightarrow \infty$ expansion
	\be
	\overline{|R_{nm}|^2|R_{pq}|^2} -  \overline{|R_{nm}|^2}\,\overline{|R_{pq}|^2} = \delta_{np}\delta_{mq}\;.
	\ee
	
	This provides of a connected contribution to \eqref{eq:product2pt}, sensitive to the variance of the matrix elements of the ensemble of operators. In the large-$N$ limit, this quantity admits a saddlepoint approximation
	\be
	\overline{G_\beta(\tau) G_{\beta'}(\tau')} \Big|_{c}\,  \sim  \, e^{-(\beta-\tau) E - \tau E' -(\beta'-\tau') E - \tau E' + S(E) + S(E')- 2f(E,E') +\beta F(\beta)+\beta'F(\beta')}\;,\label{eq:twoptwhaction}
	\ee
	for the microcanonical energies solving the system
	\begin{gather}
		\beta_{E} = (\beta-\tau) + (\beta' - \tau') + \Delta \tau_-\,\label{eq:twoptwhme}\;,\\[.4cm]
		\beta_{E'} = \tau + \tau' + \Delta \tau_+\;.\label{eq:twoptwhme2}
	\end{gather}

\begin{figure}[h]
		\centering
		\includegraphics[width=.46\textwidth]{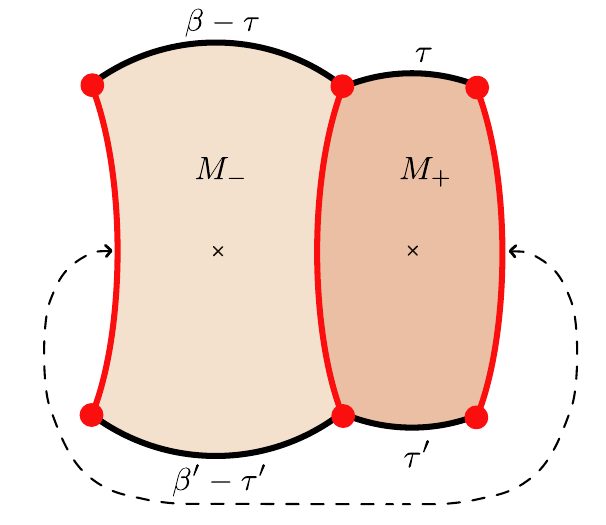}
		\hspace{1.5cm}
		\includegraphics[width=.25\textwidth]{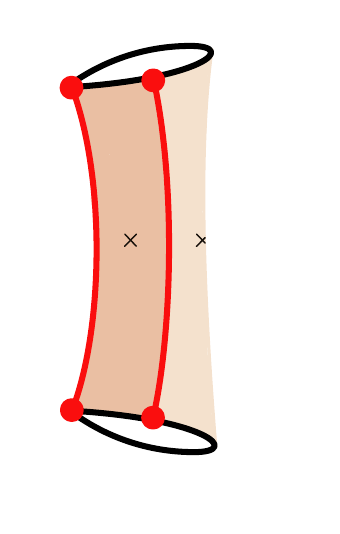}
		\caption{The two-point wormhole is deconstructed on the left figure. It corresponds to a pair of Euclidean black holes, glued together by the trajectories of two thin shells. On the right, the resulting $\mathbf{R}\times \mathbf{S}^1\times \mathbf{S}^{d-1}$ topology connecting both boundaries.}
		\label{fig:2ptwh}
	\end{figure}

	The construction of the bulk wormhole that provides this contribution is represented in Fig. \ref{fig:2ptwh}. The wormhole is built out of a pair of Euclidean black holes of masses $M\pm$, glued together along the trajectories of two thin shells which terminate at the operator insertions of the respective asymptotic boundary. The microcanonical saddlepoint equations which determine the pair $M_+,M_-$ are given precisely by \eqref{eq:twoptwhme} and \eqref{eq:twoptwhme2}, and the on-shell action of the wormhole coincides with \eqref{eq:twoptwhaction}.
	
	A similar construction follows for the connected part of the product of any two $n$-point functions. The construction generally involves a set of different wormhole solutions, related by cyclic permutations of the insertions on one of the two boundaries. 
	
\begin{figure}[h]
		\centering
		\includegraphics[width=.45\textwidth]{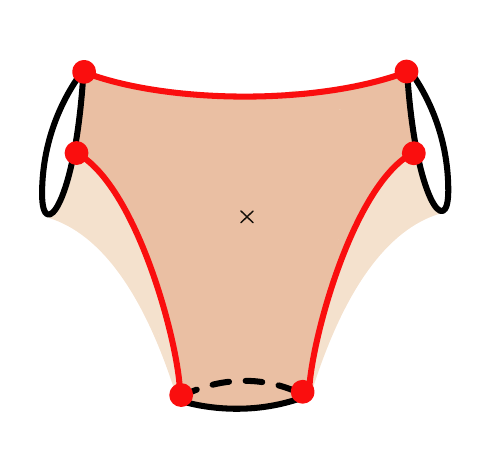}
		\caption{Three-boundary wormhole $X^{(3)}_{\text{bh}}$ constructed by gluing two Euclidean black holes of masses $M_\pm$ along the trajectories of three thin shells.}
		\label{fig:3bdywh}
	\end{figure}

	Multi-boundary wormholes arise in connected contributions to products of a larger number of correlation functions, as long as all of the $\mathcal{O}$ insertions can be paired with respective $\mathcal{O}^\dagger$ insertions. For instance, the simplest case is to consider the triple product of two-point functions. This contribution is reproduced by a three-boundary wormhole
	\be
	\overline{G_{\beta}(\tau)G_{\beta}(\tau)G_{\beta}(\tau)}\Big|_{c} \sim e^{-\Delta I[X^{(3)}_{\text{bh}}]}\,,
	\ee
	which is represented in Fig. \ref{fig:3bdywh}.

	It follows from the classical Feynman rules of the shell operator that the ETH coefficients $R_{nm}$ will satisfy Gaussian statistics at leading order in the large-$N$ expansion
	\be\label{eq:statisticsETH}
	\overline{R_{nm}R_{pq}R^*_{n'm'}R^*_{p'q'}} \approx \delta_{nn'}\delta_{mm'}\delta_{pp'}\delta_{qq'} + \delta_{np'}\delta_{mq'}\delta_{pn'}\delta_{qm'}\;.
	\ee
	
	There will be subleading non-planar corrections to the statistics of the ETH ensemble, affecting \eqref{eq:statisticsETH} when $n=p$ or $m=q$. These explain the fact that, for instance, there does not exist a completely connected contribution to $\overline{\langle \mathcal{O}\rangle_\beta \langle \mathcal{O}^\dagger\rangle_\beta G_{\beta}(\tau)}\Big|_{c}$.
	
	As pointed out in \cite{Jafferis:2022uhu} (see also \cite{Foini:2018sdb}), ETH non-gaussianities are neccesary to reproduce crossing-symmetric thermal four-point correlation functions of local conformal primaries. In our case the operator $\mathcal{O}$ is completely non-local on the $\mathbf{S}^{d-1}$, so crossing-symmetry in the OPE is not manifest. In particular, there is no stable saddlepoint to the four-point function in which the two shells can cross.
	
	\section{Approximate global symmetries and wormholes}
	\label{sec:4}
	
	Euclidean wormholes stabilized by heavy particles have been identified as a source of non-perturbative violations of bulk global symmetries in two-dimensional models \cite{Hsin:2020mfa} (see also \cite{Belin:2020jxr}), in accord with the general expectation that there are no bulk global symmetries in quantum gravity \cite{Banks:2010zn,Harlow:2018tng} \footnote{See also  \cite{Chen:2020ojn} in the context of replica wormholes.} . In this section we use the wormholes constructed in the previous section, together with their boundary interpretation, to extend these conclusions into higher dimensional AdS/CFT. 
	
	Throughout the discussion, we have made the role of global symmetries somewhat implicit in the set of bulk Feynman rules assigned to the thin shell operator. As it turns out, the shell operator $\mathcal{O}$ must carry a global flavour charge in the bulk effective description associated to the creation of the dust particles. At leading order in the large-$N$ expansion, this charge is conserved along the trajectory of the shell, until the shell reaches the insertion $\mathcal{O}^\dagger$, which provides the opposite flavor charge necessary to annihilate all of the dust particles.
	
	The argument presented in \cite{Hsin:2020mfa} is that wormholes induce non-perturbative violations of this or any other kind of global symmetry that might \textit{a priori} be present in the bulk effective theory, provided that the heavy particles are charged under the symmetry. The reason is that wormholes provide a non-vanishing contribution to products of correlation functions which should otherwise vanish, such as
	\be\label{eq:product1ptglobal}
	\overline{\la \mathcal{O} \ra_{\beta} \la \mathcal{O}^\dagger \ra_{\beta}} = e^{-I[\Delta X_{\text{bh}}]}\,,
	\ee
	for $X_{\text{bh}}$ the `black wormhole' that we constructed in section \ref{sec:31}, and we are taking $\beta$ sufficiently large, so that the contribution from the `AdS wormhole' $X_{\text{AdS}}$ can be neglected.

	Of course, the natural expectation is that the flavor symmetry of the shell is explicitly broken by the interactions of the bulk effective theory, which alter the correlation functions at subleading orders in $1/N$. Similarly, for any other global symmetry, the naive expectation is that the symmetry gets explicitly broken by some operator of large dimension in the bulk effective description, suppressed by some integer power $\alpha$ of the energy in units of the Planck (or cutoff) scale $(E/M_P)^\alpha$. These perturbative effects will clearly eclipse the role of any possible wormhole contribution. 
	
	However, it is definitely interesting to note that even if the global symmetry was protected by the whole tower of higher-dimensional operators in the bulk EFT for some reason, then the contribution of the wormhole \eqref{eq:product1ptglobal} would provide the leading non-perturbative bulk amplitude to still allow to conclude that there are no exact global symmetries in the bulk. 
	
	In the previous section, we have shown that \eqref{eq:product1ptglobal} arises in the CFT as an statistical average over an ensemble of heavy operators, whose matrix elements differ in the fine-grained structural details between chaotic eigenstates. Therefore, we identified the precise sense in which \eqref{eq:product1ptglobal} is realized for the shell operator,
	\be\label{eq:product1ptglobal2}
	\la \mathcal{O} \ra_{\beta} \la \mathcal{O}^\dagger \ra_{\beta} \approx \overline{\la \mathcal{O} \ra_{\beta} \la \mathcal{O}^\dagger \ra_{\beta}} \sim  e^{-I[\Delta X_{\text{bh}}]}\,.
	\ee
	 In this form, the wormhole is able to detect the characteristic magnitude of the thermal one-point function in the ensemble of microscopic operators which are compatible with the effective thin shell description in the bulk. Viewing the original shell operator $\mathcal{O}$ as a typical member of the ensemble for this particular correlation function, then \eqref{eq:product1ptglobal} gives the magnitude of the non-perturbative global symmetry violating amplitude in the one-point function of the shell operator. In sec \ref{sec:32} we computed the action of this wormhole as a function of the temperature, and analytically showed that for $d=2$ the action it is suppressed by the thermal entropy $e^{-S(\beta)}$ at high temperatures, while it is dominated by the Casimir energy $e^{\beta E_0}$ at low temperatures (see Fig. \ref{fig:whaction}).

	Lastly, if the bulk symmetry was gauged, then the semiclassical contribution to \eqref{eq:product1ptglobal} vanishes since the wormhole is no longer a classical solution to the equations of motion for the gauge field, failing to satisfy Gauss' law. This simply represents the well-known fact that spacetime wormholes cannot serve to propagate gauge charge between different regions. In terms of the ensemble of heavy operators, the lack of a wormhole solution is consistent with the existence of an actual global symmetry in the fundamental description. Since this charge can be resolved in the bulk effective description, the ETH ensemble of heavy operators must carry the same global charge in the fundamental description \cite{Belin:2020jxr}. The one-point function must trivially vanish within the `charged ETH ensemble' of heavy operators.

	\section{Conclusions and outlook}

	The concrete evidence presented in this paper supports the modern idea that the semiclassical path integral in higher dimensional AdS gravity is providing a coarse-grained characterization of the chaotic structure of the underlying holographic system. Specifically, the semiclassical bulk description is only able to retain an effective statistical description of the chaotic properties of the CFT, ultimately associated to black hole microstates.
	
	In this paper, we have assumed dynamical chaos in the CFT description to construct an ensemble of microscopic operators compatible with the a bulk effective description given in terms of a heavy thin shell of dust particles. Namely, the operators in the ensemble differ in fine-grained structural details of the matrix elements in the energy basis. The bulk thermal two-point function only depends on the typical magnitude of these matrix elements, which is controlled by the smooth envelope function shared by all of the operators of the ensemble. This function modifies the large-$N$ microcanonical saddlepoint contribution of the thermal correlation function, and this is dual to the backreaction of the thin shell on the spacetime geometry.
 
    Crucially, we have found semiclassical wormhole solutions which reproduce the statistical variances and higher cumulants of the correlation functions over this ensemble of operators. This allows to conclude that the semiclassical description of the thin shell is equivalent to a statistical characterization of the dual microscopic operator, namely in terms of the ensemble of operators that we have defined.
    
    Finally, we have briefly explored the role of these wormholes in the context of global symmetry violating amplitudes in higher dimensional AdS/CFT, and argued that the classical action of the one-point wormhole provides the characteristic magnitude of a non-perturbative global symmetry violating amplitude in the bulk.
	
	Our discussion in this paper has been restricted to semiclassical correlation functions. However, it is well known the semiclassical path integral can also be used to prepare gravitational states, cutting the Euclidean section open along an axis of time-reflection symmetry in the bulk. If we cut the two-point correlation function of section \ref{sec:2}, taking $\tau = \beta_L$ and $\beta - \tau = \beta_R$, we will prepare a two-sided CFT state of the form
	\be\label{eq:pets}
	\ket{\Psi} \,\propto\,\sum_{n,m} e^{- \beta_L E_n -\beta_R E_m}\,\mathcal{O}_{nm} \ket{n}\otimes \ket{m}\;,
	\ee 
	providing a realization of the `partially entangled thermal states' of \cite{Goel:2018ubv} in holographic CFTs (see also \cite{Kourkoulou:2017zaj} for one-sided analogs). From this point of view, some of the wormholes constructed in this paper, in particular those which compute overlaps between thin shell states, can be viewed as the two-sided generalizations of the wormholes presented in \cite{Chandra:2022fwi}. Also from this point of view, there is a natural relation between the ETH coarse-graining for the shell operator analyzed in this paper and the coarse-grained thermal description of the density matrix $\rho_R = \text{Tr}_L \ket{\Psi}\bra{\Psi} \approx \rho_{\beta_R}$ for the state \eqref{eq:pets}. Holographically, this is understood from the fact that the geometry outside of the black hole is agnostic about the presence of the shell in the black hole interior \cite{Engelhardt:2018kcs,Chandra:2022fwi}. 
	
	In this direction, it would be interesting to study the reconstruction properties of the state \eqref{eq:pets} in the computational basis of the black hole interior, in particular to relate the form of \eqref{eq:pets} to the non-isometric nature of the interior reconstruction \cite{Balasubramanian:2022fiy,Akers:2022qdl}. For this purpose, it would be technically convenient to consider the state \eqref{eq:pets} for a two-dimensional holographic CFT, which allows to have more control over the precise coefficients of the thin shell operator \cite{Anous:2016kss,Cardy:2017qhl}.
	
	An alternative research avenue is to analyze the role of wormholes in the real-time dynamics of thin shell operator, by taking the analytic continuation of the Euclidean correlation functions into Lorentzian time. For instance, it is straightforward to see that under the ergodic time-evolution of the CFT Hamiltonian $H$, the noise amplitude in the `operator spectral form factor' is given by the one-point wormhole that we found
	\be
	\lim\limits_{T\rightarrow \infty}\,\dfrac{1}{T}\int_{0}^T \text{d}t \;\la \mathcal{O}\ra_{\beta+it}\la \mathcal{O}^\dagger\ra_{\beta-it} \sim e^{-\Delta I[X_{\text{bh}}]}\;.
	\ee
	In this sense, it would be interesting to study dynamical probes of information loss in black hole background, such as the two-point function of the thin shell, and see whether wormholes can provide the details of the quantum noise associated to the dominant black hole microcanonical band \cite{Maldacena:2001kr,Barbon:2003aq,Barbon:2014rma,Stanford}, in particular when it comes to the characteristic timescale between $O(1)$ peaks of this noise. We leave these problems for future investigation.

	\vspace{.3cm}

	\noindent{\bf Acknowledgments}  
	
	I want to thank Vijay Balasubramanian, Jos\'{e} Barb\'{o}n, Jan de Boer, Anatoly Dymarsky, Eduardo Garc\'{i}a-Valdecasas, Tom Hartman, Albion Lawrence, Diego Liska, Javier Magán and  Boris Post for stimulating conversations. I am supported by the U.S. Department of Energy grants DE-SC0009986 and DE-SC0020360 under the HEP-QIS QuantISED program.

	\appendix
	\section{Details of the bulk two-point function}
	\label{appendix:A}
	
	In this appendix, we provide a detailed analysis of the bulk two-point function of the thin shell operator, whose general features are presented in section \ref{sec:22}.
	
	\subsection{Thin shell formalism}
	
	Let us start with providing a quick review of the details of the thin-shell formalism. Consider the general problem of gluing two Euclidean Schwarzshild-AdS regions $X^\pm$ with local geometry
	\be\label{eq:appAgeometry}
	\text{d}s_\pm^2\, = \,  f_\pm(r)\,\text{d}\tau_\pm \,+\,\dfrac{\text{d}r^2}{f_\pm(r)}\,+\,r^2\,\text{d}	 \Omega_{d-1}^2\;,
	\ee
	which could as well be considered Euclidean AdS if either one of them corresponded to the $M_\pm=0$ solution.

	The gluing of both Euclidean spaces $X^\pm$ along the trajectory $\mathcal{W}$ of a localized codimension-one domain wall must be performed according to Israel's junction conditions (cf. \cite{poisson2004relativist}). Consider the pair $(h^\pm_{ab},K_{ab}^\pm)$ of induced metrics $h^\pm_{ab}$ and extrinsic curvatures $K_{ab}^\pm$ evaluated at $\mathcal{W}$, in terms of the metric on each side $X^\pm$.  Defining $\Delta h_{ab} = h_{ab}^+ - h_{ab}^-$ and $\Delta K_{ab} = K_{ab}^+ - K_{ab}^-$, the junction conditions simply state that
	\begin{gather}
	\Delta h_{ab} =0 \,,\label{junc1}\\[.4cm]
	\Delta K_{ab}- h_{ab} \Delta K = -8\pi G S_{ab}\,,\label{junc2}
	\end{gather}
	where $\Delta K  = h^{ab}\Delta K_{ab}$.\footnote{There is an overall minus sign in this expression with respect to the Lorentzian version, from the fact that the hypersurface $\mathcal{W}$ is spacelike.} The quantity $S_{ab}$ is the localized energy-momentum of the thin domain wall.  For a thin shell of dust particles, we have $S_{ab} = -\sigma u_au_b$. \footnote{The minus sign in $S_{ab}$ comes from the analytic continuation $\sigma \rightarrow -\sigma$ from the Lorentzian perfect fluid.}
	
	The junction conditions determine the equations of motion for the thin shell, governing the trajectory $R=R(T)$, for $T$ the synchronous proper time of the particles that form it. First of all, it follows that the angular parts of \eqref{junc2} impose that the parameter
	\be
	m = \sigma V_\Omega R^{d-1}\,
	\ee
	is conserved along $\mathcal{W}$. This quantity has the obvious interpretation of the rest mast of the thin shell.
	
	The continuity of the metric \eqref{junc1} relates
	\be\label{eq:appextrinsic}
	f_\pm \dot{\tau}_\pm = \sqrt{-\dot{R}^2 + f_\pm} \,,
	\ee	
	where $\dot{x} = dx/dT$ and the square root can have either sign, depending on the particular trajectory of the shell.
	
	Finally, the remaining component of \eqref{junc2} imposes that
	\be\label{eq:appjcext}
	\kappa_+ \sqrt{-\dot{R}^2+f_+(R)} - \kappa_- \sqrt{-\dot{R}^2+f_-(R)} = \dfrac{8\pi G m}{(d-1)V_\Omega R^{d-2}}\,.
	\ee
	Here $\kappa_\pm = \text{sign}(\dot{\tau}_\pm)$ is the sign of the extrinsic curvature. Squaring this expression, one gets to the effective equation of motion for a non-relativistic particle of zero total energy
	\be\label{eq:appeom}
	\dot{R}^2 + V_{\text{eff}}(R) = 0\;,
	\ee
	subject to the effective potential
	\be\label{eq:appeffpotential}
	V_{\text{eff}}(R) = -f_+(R) + \left(\dfrac{M_+-M_-}{m} - \dfrac{4\pi G m}{(d-1)V_\Omega R^{d-2}}\right)^2\;.
	\ee
	The shell starts at $R=r_\infty$ and has a turning point at $R = R_*$ with $V(R_*)=0$. It is easy to see that $R_*\geq r_\pm$ for the respective horizon radii.
	
	The Euclidean time elapsed by the shell can be computed from \eqref{eq:appextrinsic} and it is given by the integral
	\be\label{eq:appetime}
	\Delta \tau_\pm \, = \, 2\,\int_{R_*}^\infty \dfrac{\text{d}R}{f_\pm}\,\sqrt{\dfrac{f_\pm + {V}_{\text{eff}}}{-{V}_{\text{eff}}}}\;.
	\ee

	\subsubsection*{2+1 dimensions}
	
	For $d=2$, the above expressions can be evaluated more explicitly. The BTZ horizon radii are given by $r_\pm = \sqrt{8GM_\pm}$, and the inverse temperatures by $\beta_\pm     = 2\pi/r_\pm$. The effective potential \eqref{eq:appeffpotential} in this case is quadratic
	\be\label{2+1potential}
	{V}_{\text{eff}}(R)\, = \, -(r^2-R_*^2)\;,
	\ee  
	for the turning point
	\be\label{2+1radius}
	R_*\, = \, \sqrt{r_+^2\,+\,\left(\dfrac{M_+-M_-}{m}\,-\,2Gm\right)^2}\;.
	\ee
	
	The solution of the shell's trajectory \eqref{eq:appeom} can be explicitly found
	\be\label{eq:app2+1sol}
	R(T)\, = \, R_*\,\cosh T\;,
	\ee
	where we chose the initial condition such that the shell passes through $R_*$ at proper time $T=0$.
	
	The Euclidean time elapsed by the shell \eqref{eq:appetime} can also be computed analytically. It gives
	\be\label{etime2+1}
	\Delta \tau_\pm \, = \beta_\pm \,\dfrac{\arcsin(r_\pm/R_*)}{\pi}\;.
	\ee

	\subsection{Structure of the bulk two-point function}
	
\begin{figure}[h]
		\centering
		\includegraphics[width=.385\textwidth]{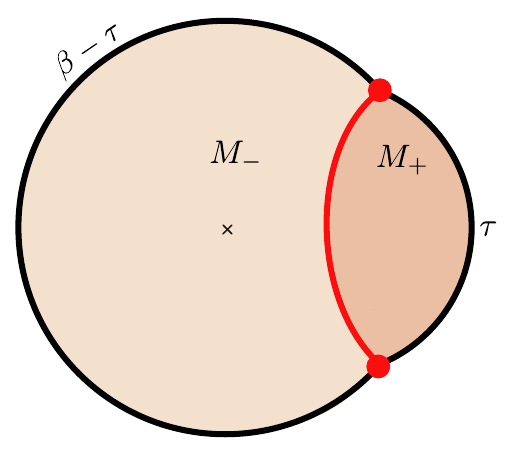}
		\caption{Saddlepoint manifold $X$ for $\tau <\tau_c$.}
		\label{fig:bulk2ptapp}
	\end{figure}

	Let us consider the structure of the gravitational saddlepoint $X$ for the bulk two-point function, without loss of generality for $M_+ \geq M_-$, corresponding to $\tau \leq \frac{\beta}{2}$. \footnote{The complementary regime is obtained from the continuation $\tau \rightarrow \beta -\tau$.} For small values of $\tau$, the solution is shown in Fig. \ref{fig:bulk2ptapp}. It will satisfy
	\be\label{eq:outsideregimeappendix}
	M_+-M_- > \dfrac{4\pi G m^2}{(d-1)V_\Omega R_*^{d-2}}\,.
	\ee
	In terms of state preparation along the axis of time-reflection symmetry, this is simply the statement that the shell sits outside of the right horizon, and the right ADM mass must increase by at least the self-energy of the shell at the turning point. The relevant signs of the extrinsic curvatures in the Euclidean junction condition \eqref{eq:appjcext} are all positive in this case, $\kappa_\pm =1$.
	
	The Euclidean microcanonical saddlepoint equations in this case read 
	\begin{gather}
	\beta_- = \beta- \tau + \Delta \tau_-\;,\label{eq:bulksaddle1ap}\\[.2cm]
	\tau = \Delta \tau_+\,.\label{eq:bulksaddle2ap}\,
	\end{gather}

	However, as $\tau$ gets increased, $M_+-M_-$ will decrease, from the fact that the shell passes closer to the right horizon. For some value of $\tau =\tau_c$, it will be the case that \eqref{eq:outsideregimeappendix} gets saturated. When this happens, we have $V_{\text{eff}}(R_*) = -f_+(R_*) = 0$ so that $R_* = r_+$, and the shell bounces back precisely at the right horizon. The energy difference is therefore fixed in this case
	\be\label{eq:outsideregimeappendix}
	M_+-M_- = \dfrac{4\pi G m^2}{(d-1)V_\Omega r_+^{d-2}}\,.
	\ee
	For $d=2$ this gives $M_+-M_- = Gm^2$, independent of the total energy $\bar{M} = \frac{M_++M_-}{2}$. In higher dimensions, this will generate a relation of the form $\omega = \frac{M_+-M_-}{2} = \omega_c(\bar{M})$. \footnote{Since the right hand side of \eqref{eq:outsideregimeappendix} is a monotonically decreasing function of $M_+-M_-$, there will be only a solution to this equation for a fixed value of $\bar{M}$. } So, we see that the previous solution will only exist for $\omega \geq \omega_c$. The value of the critical Euclidean time $\tau_c= \tau_c(\beta)$ solves the system \eqref{eq:bulksaddle1ap} and \eqref{eq:bulksaddle2ap} together with the extra condition \eqref{eq:outsideregimeappendix}.
	
	\subsubsection*{Fitting the envelope function for $\tau < \tau_c$}
	
	Under the identification $E= M_-$ and $E'= M_+$, the microcanonical saddlepoint equations \eqref{eq:saddle1} and \eqref{eq:saddle2} reduce to the bulk system \eqref{eq:bulksaddle1ap} and \eqref{eq:bulksaddle2ap} in the regime $\omega \geq \omega_c(\bar{E})$, provided that
	\begin{gather}
	\Delta \tau_- = \partial_- f \;,\label{eq:bdycondoutapp}\\[.2cm]
	\beta_+ - \Delta \tau_+= \partial_+ f\label{eq:bdycond2outapp}\,
	\end{gather}
	It is possible to check from the general expressions \eqref{eq:shifttime} that 
	\be
	\partial_+ \Delta \tau_- = - \partial_- \Delta \tau_+\,
	\ee
	in this regime, so that the solution for $f(E,E')$ also exists. \footnote{The difference in sign with the case analyzed in section \ref{sec:2} arises from the fact that the square root in \eqref{eq:shifttime} selects the opposite branch in this regime.}
	
	To get the value of the function, we shall evaluate the on-shell action $I[X]$, by adding and subtracting the part associated to the region $X_s^+$ which is now absent in the Euclidean solution (right part of the green region in Fig. \ref{fig:bulk2ptaction}).  The total action reads
	\be
	I[X] \,=\, (\beta_--\Delta\tau_-) F(\beta_-) + \Delta\tau_+ F(\beta_+) + I[X'_s]\;,
	\ee
	where 
	\be
	I[X_s'] = \dfrac{d}{8\pi G}\,\left[\text{Vol}(X_s^-)- \text{Vol}(X_s^+)\right]\,+\,m\dfrac{d-2}{d-1}  L[\gamma_{\mathcal{W}}]  \;,
	\ee
	in terms of the quantities defined in \eqref{eq:length} and \eqref{eq:vol}. 
	
	Comparing the on-shell actions, the value of the envelope function for $\omega \geq \omega_c$ must then be
	\be
	f(E,E') = \alpha_- S(E) + \alpha_+ S(E') + I[X_s']\;,
	\ee
	for $\alpha_\pm = \frac{\Delta \tau_\pm}{\beta_\pm}$. It is easy to see that $I[X_s']$ is the analytic continuation of $I[X_s]$ in \eqref{eq:Ishell2}. The relative minus sign in the volume difference in $I[X_s']$ cancels with the opposite branch selected by the square root in \eqref{eq:vol}.
	
	\subsubsection*{Structure of the solution at $\tau > \tau_c$}
	
	Lastly, for Euclidean times larger than the critical time $\tau > \tau_c(\beta)$, the trajectory of the shell moves into the opposite part of the Euclidean disk, and the situation is the one analyzed in section \ref{sec:2} (see Fig. \ref{fig:bulk2ptaction}). However, there is a subtlety which has been omitted in the analysis of section \ref{sec:2} for the sake of clarity of the argument, since the subtlety does not alter any of the results. The subtlety is the fact that, for $d>2$, the trajectory of the shell must curve back at some large radius $r_\star > R_*$. This radius is defined by the condition	
	\be\label{eq:outsideregimeappendix}
	M_+-M_- = \dfrac{4\pi G m^2}{(d-1)V_\Omega r_\star^{d-2}}\,.
	\ee
	
	So, for $R_* \leq R \leq r_\star$, the sign of the extrinsic curvature in the junction condition \eqref{eq:appjcext} is, $\kappa_\pm =\mp 1$, while for $R>  r_\star$ the right sign switches to $\kappa_\pm = 1$.
	
	This subtlety affects the value of $\Delta \tau_+$, in particular, it decreases it slightly, since there is a portion of the trajectory, close to the asymptotic boundary, for which $\dot{\tau}_+$ changes sign. However, this value will still be given by the formal expression \eqref{eq:shifttime}, for which this change in sign is implicit in the choice of branch of the square root in the numerator. Thence, none of the results presented in section \ref{sec:2} is affected by this subtletly.

	\bibliographystyle{ourbst}
	\bibliography{wh.bib}
	
\end{document}